\documentclass[sigchi,nonacm,manuscript]{acmart}

\usepackage{listings}
\usepackage{multirow}

\usepackage{subcaption}
\AtBeginDocument{%
  }

\setcopyright{acmlicensed}
\copyrightyear{2025}
\acmYear{2025}
\acmDOI{XXXXXXX.XXXXXXX}
\acmConference[Conference acronym 'XX]{Make sure to enter the correct
  conference title from your rights confirmation email}{June 03--05,
  2018}{Woodstock, NY}
\acmISBN{978-1-4503-XXXX-X/2018/06}

\begin{document}

\title{MimiTalk: Revolutionizing Qualitative Research with Dual-Agent AI}

\author{Fengming Liu}
\email{leo.liu.23@ucl.ac.uk}
\orcid{0009-0009-3881-496X}
\affiliation{%
  \institution{University College London}
  \city{London}
  \country{United Kingdom}
}

\author{Shubin Yu}
\email{yu@hec.fr}
\orcid{0000-0001-7719-3056}
\affiliation{
  \institution{HEC Paris}
  \city{Paris}
  \country{France}
}

\authornote{All authors contributed to the conceptualization and idealization of this research. F.L. led the data analysis and writing of the manuscript. S.Y. provided supervision and guidance throughout the research process. Both authors reviewed and approved the final manuscript.}

\renewcommand{\shortauthors}{Liu and Yu}

\begin{abstract}
We present MimiTalk, a dual-agent constitutional AI framework designed for scalable and ethical conversational data collection in social science research. The framework integrates a supervisor model for strategic oversight and a conversational model for question generation. We conducted three studies: Study 1 evaluated usability with 20 participants; Study 2 compared 121 AI interviews to 1,271 human interviews from the MediaSum dataset \citep{zhuMediaSumLargescaleMedia2021} using NLP metrics and propensity score matching; Study 3 involved 10 interdisciplinary researchers conducting both human and AI interviews, followed by blind thematic analysis. Results across studies indicate that MimiTalk reduces interview anxiety, maintains conversational coherence, and outperforms human interviews in information richness, coherence, and stability. AI interviews elicit technical insights and candid views on sensitive topics, while human interviews better capture cultural and emotional nuances. These findings suggest that dual-agent constitutional AI supports effective human-AI collaboration, enabling replicable, scalable and quality-controlled qualitative research.
\end{abstract}



\begin{teaserfigure}
    \includegraphics[width=\textwidth]{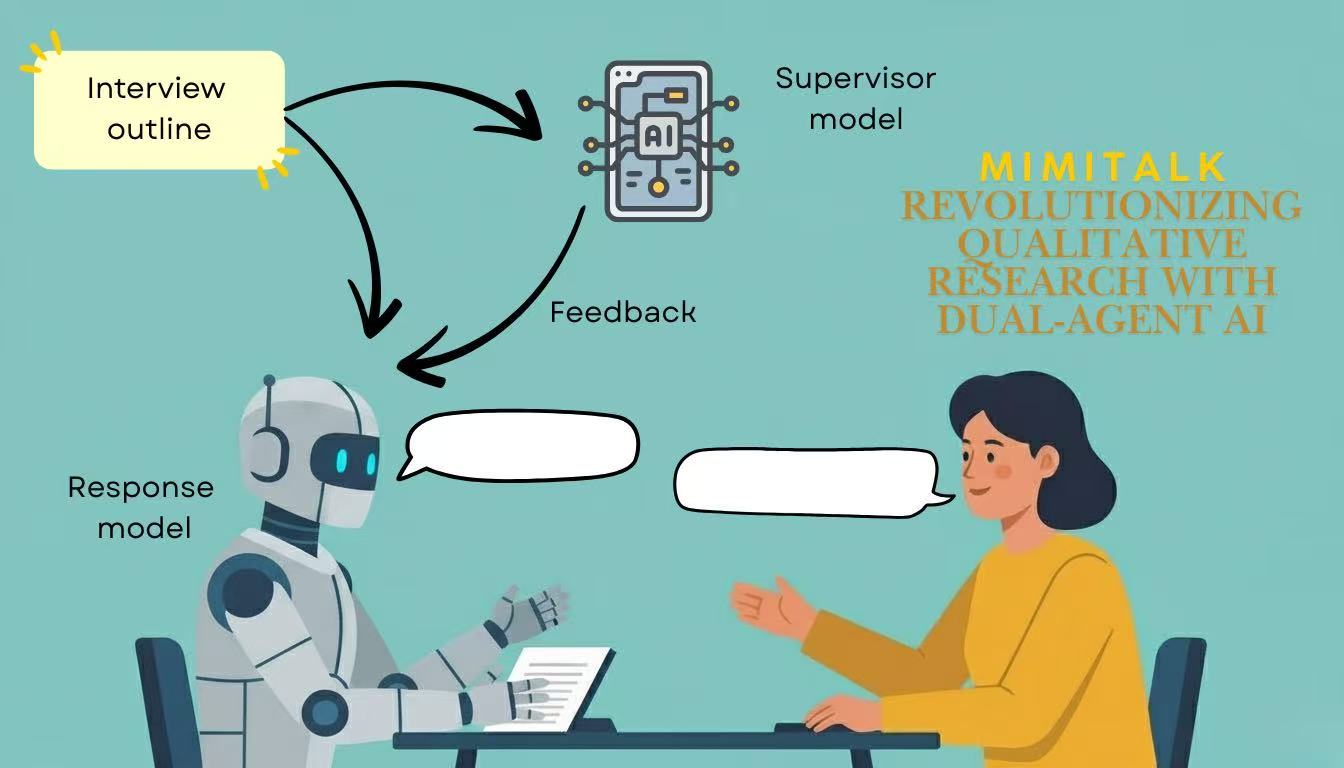}
    \caption{MimiTalk: Revolutionizing Qualitative Research with Dual-Agent AI}
    \Description{MimiTalk.app}
    \label{fig:teaser}
\end{teaserfigure}

\maketitle

\section{Introduction}
The rapid advancement of generative artificial intelligence (large language models, LLMs) has fundamentally transformed how humans interact with artificial intelligence, making increasingly sophisticated conversational exchanges blur the boundaries between human and machine communication. This technological evolution presents unprecedented opportunities for automating traditionally human-intensive tasks, particularly in research contexts where conversation-based inquiry is central to data collection. However, the potential for AI systems to effectively conduct interviews—one of humanity's most fundamental social activities—remains largely unexplored in Human-Computer Interaction (HCI) research.

Traditional interview methods in qualitative research face significant challenges that directly impact user experience and research outcomes: high resource demands limit study scale, interviewer bias affects response validity, and geographical constraints restrict participant access. While video conferencing technology has addressed some distance barriers, research indicates that anonymous digital interfaces can elicit more candid responses than human interviewers, particularly regarding sensitive topics \citep{pichardRoster2020}. This finding suggests that AI-driven interview systems might offer unique advantages in certain research contexts by reducing social desirability bias and interview anxiety.

Recent empirical studies have provided evidence for AI-assisted interview methods in specific research contexts. In a pioneering study involving 395 interviews on stock market participation, researchers found that AI-assisted interviews not only maintained high data quality but achieved greater analytical depth than traditional approaches \citep{chopraConductingQualitativeInterviews2023}. Interestingly, participants in this study showed a preference for AI interviewers over human ones, suggesting that AI systems might offer advantages in certain research contexts by reducing social desirability bias and interview anxiety.

Against this background, we present MimiTalk, a novel AI-powered interview framework designed to explore how AI systems can effectively conduct interviews with human participants in qualitative research contexts. Our work addresses fundamental HCI questions about AI-powered interviewing: How can AI systems conduct interviews that maintain the depth and authenticity of traditional human-led methods while offering significant scalability advantages? What are the optimal design patterns that leverage AI capabilities in conversational generation and strategic oversight? How can trust and ethical compliance be maintained in AI-powered research frameworks?

We systematically evaluate MimiTalk through three complementary studies. Study 1 conducted a usability evaluation with 20 participants recruited through Prolific, demonstrating the platform's effectiveness in reducing interview anxiety and maintaining conversational coherence. Study 2, based on 121 English AI interviews compared against 1,271 human interviews with distinguishable roles from the MediaSum dataset \citep{zhuMediaSumLargescaleMedia2021}, demonstrates that AI-led interviews significantly outperform human interviews in information richness, semantic coherence, and result stability using NLP metrics including DeBERTa \citep{heDeBERTaDecodingenhancedBERT2020} tokenizer-based information entropy and cross-turn semantic similarity, while maintaining conversational fluency. Study 3 recruited 10 interdisciplinary bilingual researchers to conduct both human and AI interviews using consistent outlines, followed by blind thematic/content analysis of verbatim transcripts. Results show that participants generally view AI as "efficiency enhancement" rather than "creativity replacement"; trust in literature search and citation generation is low, moderate for analysis and programming, and highest for writing assistance and formatting; AI interviews can elicit technical insights comparable to expert interviews and enhance candid expression on sensitive topics like academic integrity, while human interviews excel in capturing cultural context and emotional nuances.

The main contributions of this paper are as follows:

- Conceptual contribution: Through empirical research, we further validate and confirm GenAI's role as an ``evocative object'' in interview contexts for psychological projection and meaning construction, and elucidate the necessity and mechanisms of dual-agent constitutional collaboration in high-engagement tasks; we propose and demonstrate the key tension between "scalability and emotional subtlety."
- Methodological contribution: We propose a reusable dual-agent constitutional interview paradigm and process that decouples and embeds ethical compliance, quality control, and conversational generation without sacrificing conversational naturalness.
- Empirical contribution: Based on cross-corpus comparison of 121/1,271 interviews and human-AI comparison with 10 researchers, we provide systematic evidence for AI interviews in information richness, semantic coherence, technical insights, and candid expression, while characterizing boundary conditions in cultural and emotional subtlety, offering design principles and usage recommendations for scalable qualitative research.

Next, we will review related work, detail system design and experimental methods, report quantitative and qualitative results from both studies, and finally discuss implications for HCI methodology and human-AI collaboration research.

\section{Related Work}

\subsection{AI-Driven Dialogue Systems and Human-Computer Interaction}

AI-driven dialogue systems have undergone significant evolution from simple rule-based chatbots to complex neural network architectures. Early systems relied on basic pattern matching, while the emergence of Transformer architecture \citep{vaswani2017attention} brought revolutionary advances in contextual understanding and natural conversational flow. Modern dialogue management systems can handle complex conversational contexts \citep{brabraDialogueManagementConversational2022}, providing more sophisticated solutions for interview processes and participant interactions. The development of BERT \citep{devlin2018bert} and GPT series models \citep{brown2020language} established new benchmarks for human-computer interaction, paving the way for automated interview applications. Continuous improvements in language models \citep{borgeaudImprovingLanguageModels2022} continue to enhance AI systems' conversational capabilities.

Recent empirical studies have provided evidence for AI-assisted interviews in large-scale research. In a study involving 395 interviews on stock market participation, researchers found that AI-assisted interviews not only maintained high data quality but achieved deeper analytical results than traditional methods \citep{chopraConductingQualitativeInterviews2023}. Interestingly, participants in this study showed a preference for AI interviewers over human ones, challenging traditional assumptions about qualitative research methods.

The success of AI in interviews has extended beyond financial research. In recruitment contexts, AI systems demonstrate sophisticated conversational capabilities that enhance candidate engagement \citep{ahmadFutureRecruitmentUsing2024}. In healthcare, AI improves clinical decision-making through optimized patient interactions \citep{alowaisRevolutionizingHealthcareRole2023, krishnanArtificialIntelligenceClinical2023}. The financial services industry utilizes AI for predictive analysis and customer interactions \citep{bahooArtificialIntelligenceFinance2024}, while education leverages AI to create more personalized learning experiences \citep{sapciArtificialIntelligenceEducation2020}. These cross-domain applications demonstrate that AI systems can support diverse forms of human-computer interaction in specific contexts.

Field studies have also revealed interesting dynamics in AI interview applications. For example, research on asynchronous video interviews (AI-AVI) shows that when systems integrate perceptibility and transparency features, participants' cognitive trust increases \citep{suenHung2023}. The development of AI conversational interview systems \citep{wuttkeAIConversationalInterviewing2024} further expands the possibilities of automated interviews. In a large-scale study involving 710 simulated interviews, machine learning models successfully predicted personality traits (R² = 0.32) and interview performance (R² = 0.44), with accuracy exceeding self-report indicators \citep{koutsoumpis2024}. However, such evaluation systems still face challenges in non-evaluative applications like social science research.

The integration of AI systems in research contexts raises important questions about human-AI collaboration and trust calibration. Research shows that even in high-scoring tasks, users remain cautious about completely delegating citations, fact-checking, or interpretive judgments to AI, with a general "trust ceiling" of approximately 80\% \citep{openaiGPT4TechnicalReport2024}. This skepticism reflects ongoing concerns in the academic community about hallucination phenomena and limitations in AI contextual understanding. Research on retrieval-augmented generation \citep{shusterRetrievalAugmentationReduces2021} indicates that anchoring AI responses to factual information can reduce hallucination problems. Meanwhile, efforts to improve fairness in machine learning systems \citep{holsteinImprovingFairnessMachine2019} address important ethical considerations in AI applications.

In the field of AI-assisted qualitative data analysis, recent research has further validated the potential of AI tools in social science research. Akman et al. \citep{akmanHumanResearcherVs2025} compared human researchers with AI-supported qualitative data analysis and found that hybrid prompt design can significantly enhance the efficiency and accuracy of qualitative data analysis, providing deeper insights and more structured outputs. The study employed open, axial, and selective coding methods, demonstrating AI's effectiveness in programming education research while emphasizing the importance of carefully designed prompts and human oversight in reducing potential biases and errors in AI analysis. Hamilton et al. \citep{hamiltonExploringUseAI2023} compared human and AI-generated qualitative analysis in guaranteed income data research, finding both similarities and differences in theme identification, with human coders identifying some themes that ChatGPT did not, and vice versa. This finding suggests that AI can serve as a complementary tool for complex human-centered tasks, providing additional support for research tasks. Li et al. \citep{liComparingGPT4Human2024} compared GPT-4 with human researchers in healthcare data analysis, finding that GPT-4 can effectively identify key themes, showing moderate agreement with human analysis ($\kappa=0.401$), and while human analysis provided richer subtheme diversity, AI's consistency suggests its potential as a complementary tool for qualitative research.

\subsection{Dual-AI Agent as "evocative object"}
The question of whether artificial intelligence should be regarded as a tool for social science research or as a "conversationalist" with its own dialogical status remains highly controversial. Contemporary scholars often reduce the dialogue and cognitive experiences created by AI systems to purely instrumental use, essentially viewing them as deterministic, algorithmic mechanical processes. This perspective can be analogized to Searle's "Chinese Room" thought experiment \citep{searleMindsBrainsPrograms1980}. In this scenario, a person who doesn't understand Chinese is locked in a room, mechanically matching and outputting Chinese symbols based solely on an operation manual. To external observers, it appears that this person understands Chinese, but in fact, they have no real understanding. If we believe that AI systems lack true understanding or consciousness, merely manipulating symbols according to preset rules without grasping their meaning, then we are operating within the framework of the "Chinese Room."

However, despite the "Chinese Room" emphasizing AI's lack of true understanding, Dennett's "intentional stance" theory \citep{dennettIntentionalStance1998} provides a perspective more aligned with our research context. Dennett argues that when we adopt an "intentional stance" toward a system, we interpret its behavior as that of a rational agent, assuming it has beliefs, desires, and intentions. This strategy is often very effective in predicting and understanding system behavior. In our research context, this theoretical perspective helps explain why participants can engage in meaningful dialogue with AI even when they know that AI lacks true consciousness.

Furthermore, subjectivity need not be understood as an entity inherent in consciousness, but can be viewed as a position within the symbolic order. Lacan's psychoanalytic theory \citep{lacanEcritsFirstComplete2006} indicates that subjectivity is constructed through language and symbolic systems, where the subject does not exist as a transcendental autonomous entity but occupies specific symbolic positions. Žižek further argues that the core of ideological operation lies in the subject's identification with and projection onto symbolic authority, maintaining belief even when the subject knows of its fictionality \citep{zizekSublimeObjectIdeology2002}. These theories reveal a persistent impression among AI scientists and algorithm engineers: that there is a fundamental difference between human intelligence and the mechanical, programmatic nature of machines. However, in human-computer interaction contexts, if humans cannot infer understanding from AI's input and output, then they also cannot determine whether other humans truly possess such understanding. The only difference is that humans can base empathy on analogies to their own intelligence, and this analogy cannot naturally extend to AI.

Moreover, the advanced natural language processing capabilities of contemporary AI systems make their outputs no longer simply predictable. This is fundamentally different from text correction algorithms like Grammarly or Word: when humans interact with AI, they often expect results that exceed their own predictions, even hoping that AI can correct or expand their own thinking.

This phenomenon forms an interesting contrast with the findings of Luger et al. \citep{lugerHavingReallyBad2016}. Their research shows that users often view conversational agents as anthropomorphic assistants with subjectivity, but actual experiences are often disappointing, revealing gaps between expectations and reality.

This also explains why participants, even knowing its artificial nature, tend to attribute real subjectivity to our collaborative framework. This phenomenon reflects the psychological mechanism of humans' tendency to project social rules and expectations onto computers, as revealed by media equation research \citep{nassMachinesMindlessnessSocial2000}. We believe that MimiTalk's collaborative framework functionally resembles what Turkle describes as an "evocative object" \citep{turkleAloneTogetherWhy2011}—a technological entity capable of stimulating psychological projection and meaning construction. Due to its lack of human judgment, emotional responses, and social expectations, it instead creates a psychological space where participants can express themselves more authentically, while maintaining research quality under the premise of human researchers providing strategic supervision and contextual understanding.

\subsection{Applications of Human-Computer Collaboration in Research Contexts}

The evolution of interview methodology has been significantly influenced by technological advances, particularly in the field of human-computer interaction. Traditional interview methods are generally divided into three categories: structured interviews with fixed questions, semi-structured interviews allowing some flexibility, and unstructured interviews emphasizing open dialogue \citep{cuevas2024}. These methods have gradually evolved from simple conversations to complex approaches integrating digital interfaces and multimodal interactions.

Recent HCI research has explored the impact of digital interfaces on interview dynamics. Studies show that compared to human-conducted interviews, anonymous digital interfaces can elicit more candid responses from participants when dealing with sensitive topics \citep{pichardRoster2020}. This finding suggests that AI-driven interview systems may have unique advantages in certain research contexts, reducing social desirability bias and interview anxiety.

The theoretical framework of digital interviews has evolved from standardized protocols in the early 20th century to contemporary approaches with greater cultural sensitivity \citep{brinkmann2018}. The formulation of questions plays a crucial role in knowledge acquisition, similar to how different lenses affect photographic effects \citep{briggs2003questionformulation}. Various interview methods serve different purposes: structured interviews ensure standardization, semi-structured interviews balance flexibility with systematic inquiry, while unstructured interviews encourage free narration \citep{kvale2021, seidman2019}. The concept of "active interviewing" \citep{holsteinActiveInterview1995} further deepens our understanding of how interviewers actively participate to gain deeper insights. Research on computer-automated interviews \citep{pickardUsingComputerAutomated2020} has also demonstrated the feasibility and effectiveness of automated systems in conducting structured interviews. Comparative studies of online and offline interviews \citep{shapkaOnlineInpersonInterviews2016} reveal trade-offs between different interview modes.

However, traditional interview methods still face some challenges that AI-driven systems may provide solutions for: for example, high resource consumption severely limits research scale \citep{seidman2019}, interviewer bias affects response validity \citep{kvale2021}, and geographical constraints hinder participant accessibility. Although video conferencing technology has alleviated distance barriers, research also indicates that it may weaken the quality of interpersonal interactions \citep{brinkmann2018}.

\section{AI-Based Interview System Design}

\subsection{Platform Architecture}
We built the application platform Mimitalk.app for the MimiTalk framework, providing it for researchers and interviewees to use. MimiTalk.app adopts a modular architecture consisting of three core components: frontend interview module, backend API service, and AI inference module. The frontend is implemented based on React.js and Material UI components, providing intuitive interaction interfaces and smooth user experience for interviewers and interviewees. The backend API service is implemented based on FastAPI, providing RESTful interfaces for interview management and real-time communication based on WebSocket.

We adopted an approach where the AI interviewer actively initiates conversations, followed by users actively starting recording and using the Whisper speech model for text conversion during interviews. We believe that using AI for interviews is a novel and unique qualitative data collection method, rather than a pure simulation and replacement of human interviews. To control the immersion generated by task avatars, the MimiTalk.app interview interface does not include human-like avatars and allows users to view interviewer questions and real-time interview transcripts in text form. To reduce users' cognitive burden, the MimiTalk.app interview interface adopts a simple minimalist design, retaining only the recording button for starting responses and the button for ending interviews.

\begin{figure}[t]
  \centering
  \begin{subfigure}[b]{0.48\textwidth}
    \centering
    \includegraphics[width=\linewidth]{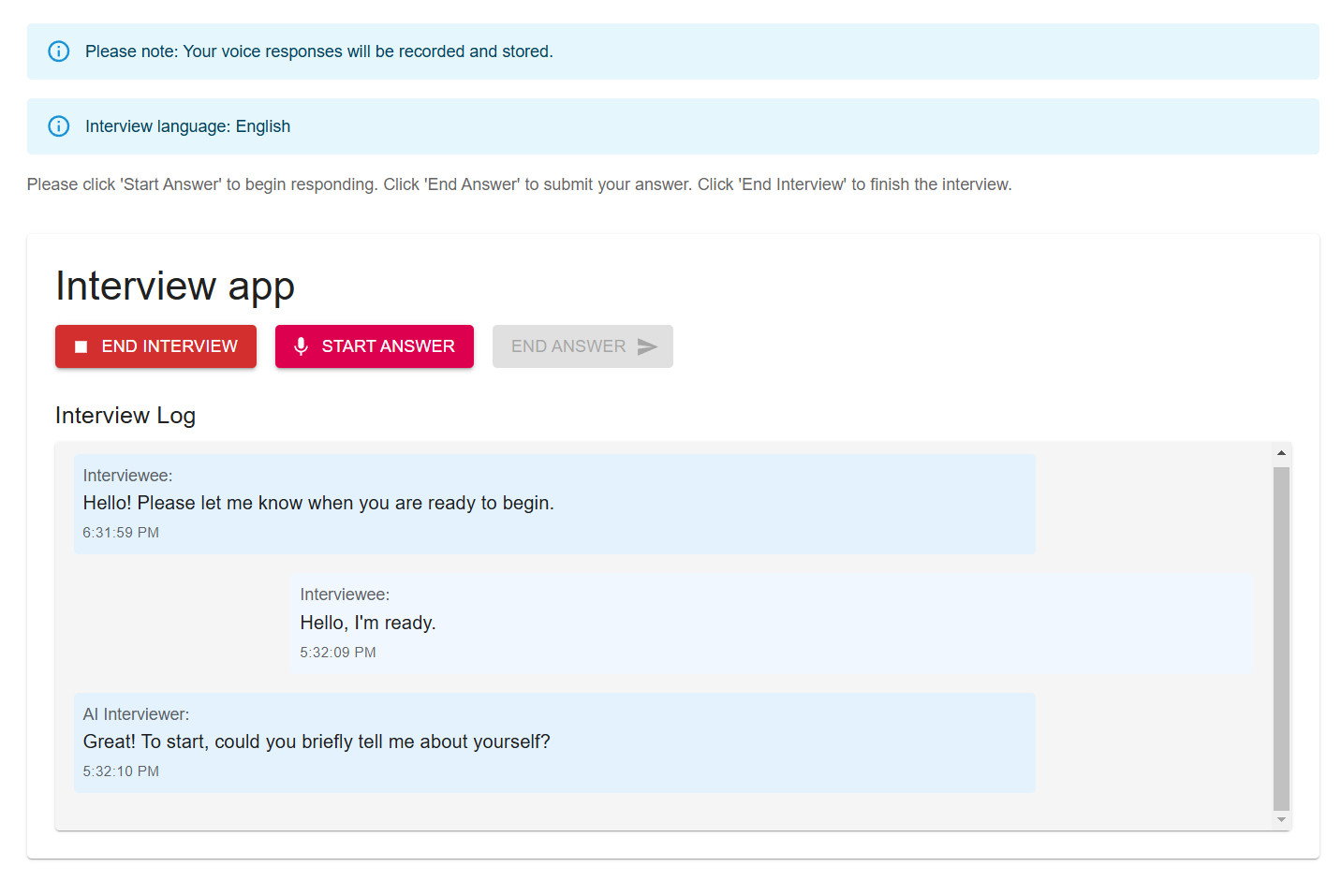}
    \caption{Screenshot of the MimiTalk.app interview interface. The interface adopts a minimalist design, retaining only core function buttons such as recording and ending the interview, avoiding anthropomorphic avatars and highlighting the conversation content.}
    \label{fig:screenshot1}
  \end{subfigure}
  \hfill
  \begin{subfigure}[b]{0.48\textwidth}
    \centering
    \includegraphics[width=\linewidth]{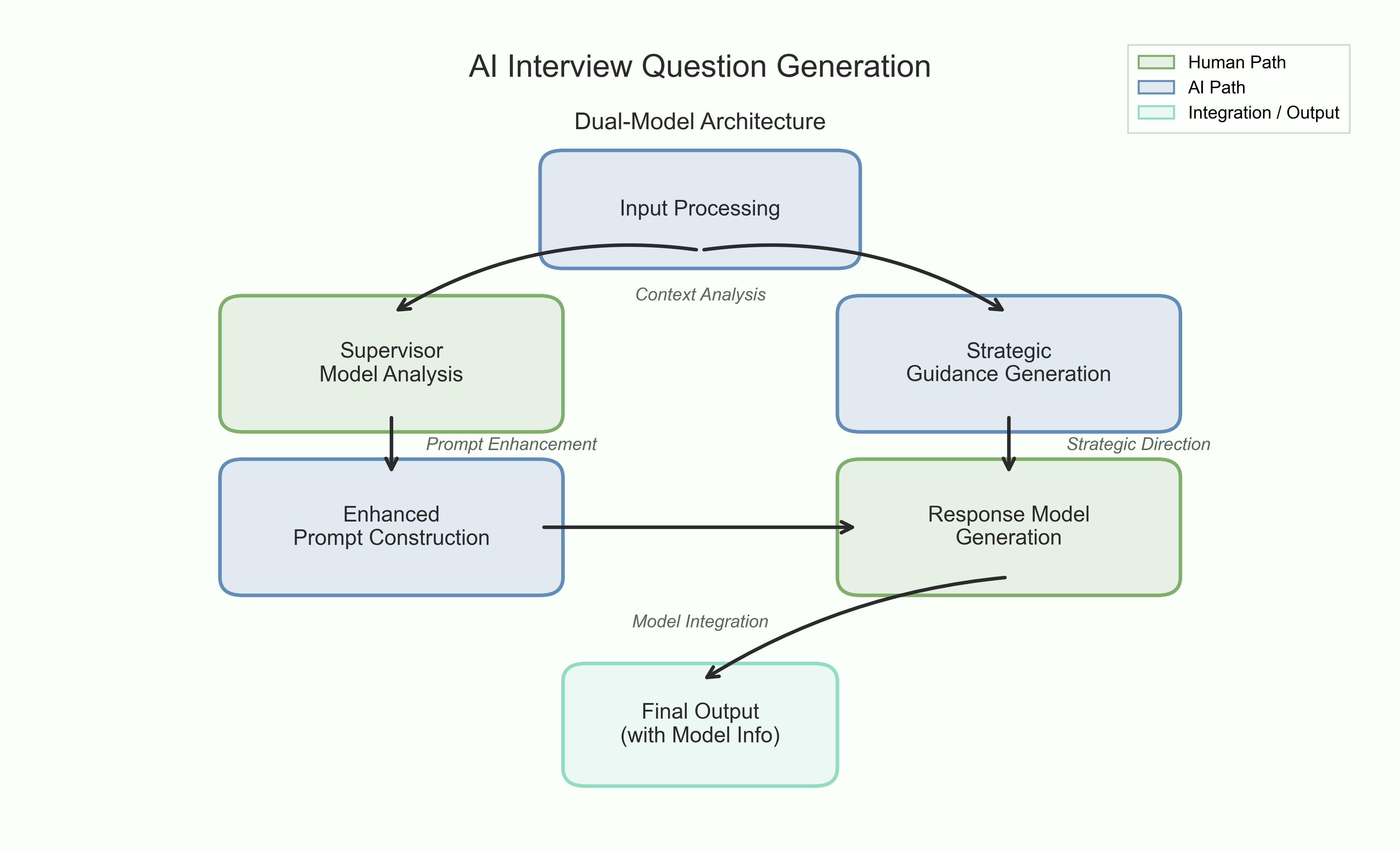}
    \caption{MimiTalk dual-agent collaborative interview system architecture. The process covers interview initialization, question generation, and response handling. The supervisor model is responsible for strategic control, while the main model generates natural conversation, ensuring compliance and fluency.}
    \label{fig:ai_logic_flow}
  \end{subfigure}
  \caption{Side-by-side display of the MimiTalk.app interview interface and system architecture: (a) The interview interface features a minimalist design to avoid confounding variables and focus on interview content; (b) The dual-agent collaborative architecture integrates constitutional principles and real-time context analysis.}
\end{figure}

\subsection{Dual-Agent Constitutional AI Architecture}
This system implements a dual-agent architecture for constitutional collaborative GenAI, with supervisor and conversational models working together to achieve controllable and scalable research interviews while ensuring ethical compliance and conversational naturalness.

\textbf{System Architecture Core}  
The core of the system lies in embedding constitutional AI principles into the dual-agent collaborative architecture, achieving organic integration of strategic supervision and natural conversation through specialized division of labor. The system adopts a modular design that decouples supervision logic from conversational generation.

\textbf{Supervisor Agent}  
The supervisor agent uses the Claude Sonnet 4.0 model, responsible for strategic-level interview quality control and ethical compliance. It continuously monitors conversational development, assesses compliance with predefined objectives, and provides real-time directional suggestions to maintain interview depth and effectiveness while ensuring ethical compliance.

\textbf{Core Multi-Model Asynchronous Architecture}  
The system implements an asynchronous architecture where the responder agent uses GPT-5 or Claude Haiku models for natural conversational generation, with intelligent model selection and background supervision analysis:

The complete implementation of the core multi-model asynchronous architecture, including the intelligent model selection and background supervision analysis functions, is provided in Appendix A.

This asynchronous architecture enables natural conversational generation with dynamic response adaptation, maintaining conversational continuity and relevance while ensuring all generated content complies with constitutional AI principles.

Figure \ref{fig:ai_logic_flow} shows the dual-model architecture for AI interview question generation. The process covers input processing, supervisor model analysis and strategic guidance generation, prompt construction and conversational generation steps.

\section{Study 1: Usability Evaluation of Mimitalk.app}

To validate the effectiveness and user experience of our AI-powered interview platform, Study 1 conducted a comprehensive usability evaluation involving 20 participants recruited through Prolific, an online research participant recruitment platform. This study serves as a foundational assessment of the platform's technical capabilities, user interface design, and overall interview experience before proceeding to more complex comparative analyses.

The participants represented diverse demographic backgrounds and geographical locations, ensuring a representative sample for our usability study. The participant pool included individuals aged 18 to 70 years (median age early 30s), with a relatively balanced gender distribution (14 female, 12 male among known participants). Educational backgrounds varied from high school/A-Levels to Master's degrees, with Bachelor's degrees being the most common qualification. The occupational profile was notably diverse, including students, professionals (such as HR advisors, financial analysts, and teachers), healthcare workers, administrative staff, homemakers, and retired individuals. Participants were predominantly from the United Kingdom, and most reported prior interview experience, primarily in job-seeking contexts, across different formats including face-to-face, online, and telephone interviews. More information of the participants in Study 1 can be found in Appendix \ref{sec:usability_demographics}.

The interview outline was structured to assess both the platform's technical capabilities and user experience. The outline can be found in Appendix \ref{sec:usability_test_outline}.

The analysis revealed that AI-generated questions showed significant advantages over static approaches through their ability to maintain conversational coherence while exploring participant expertise. However, limitations were observed including occasional redundancy, context drift in longer conversations, and cultural sensitivity gaps where AI did not always recognize nuances in academic hierarchies. The platform's ability to generate follow-up questions based on previous responses was particularly effective, with 14 of the participants noting that the questions felt contextually appropriate and engaging. 

\begin{table}[h]
\caption{Key Themes from Usability Study Qualitative Analysis}
\label{tab:usability_themes}
\small
\begin{tabular}{@{}p{4cm}|p{8cm}@{}}
\toprule
\textbf{Theme} & \textbf{Description} \\
\midrule
User Experience & Platform usability, interface interaction, response mechanisms \\
Psychological Impact & Anxiety levels, comfort, pressure management \\
Technical Performance & Response time, accuracy, system reliability \\
Comparative Elements & Differences from traditional interviews \\
Limitations & Areas for improvement and development \\
\bottomrule
\end{tabular}
\end{table}

The usability evaluation revealed several key insights about the AI-powered interview platform's effectiveness and user experience. The system indicated 41 people clicked the interview link, with 20 participants completing the full interview process, representing a $48.8\%$ completion rate that demonstrates reasonable user engagement.

Participant feedback highlighted several strengths, including natural conversation flow, comfortable interview environment, and clear and logical question progression. The interface received particular praise for its intuitive design and straightforward navigation. Many participants emphasized the platform's quick response times and accurate speech recognition capabilities, which contributed to a smooth interview process.

The psychological aspects of the AI-driven interview process emerged as a significant theme. Participants consistently reported lower anxiety levels compared to traditional interviews, citing the absence of immediate human judgment as a key factor. The ability to pace their responses and focus on content rather than physical presentation was particularly valued by participants who reported previous interview anxiety.

When comparing the platform to traditional interview experiences, participants also noted several distinctive advantages. The consistency in question delivery and standardized evaluation approach was frequently mentioned as a positive feature. Participants particularly appreciated the elimination of appearance-based bias and reduced pressure regarding body language, allowing them to focus more on their responses' substance.

However, limitations were identified including the lack of real-time response transcription, limited ability to build personal rapport compared to human-led interviews, and occasional challenges with natural language understanding. The overall sentiment was balanced, with participants endorsing the platform's effectiveness as a preliminary screening tool and suggesting its value as a complementary tool.
\section{Methodology}

\subsection{Study 2: Quantitative Evaluation of AI-powered Interviews}

\textit{Data Collection and Privacy Considerations.} All AI interview data for this study were collected through the MimiTalk.app platform, with research permissions included in the platform's terms of service. Due to user privacy protection and ethical considerations, we cannot share the raw interview data with third parties. However, future research may consider collecting additional AI interview data separately for replication studies or larger-scale investigations and mechanism studies.

To deeply evaluate the quality of AI-driven interviews, we designed a multi-dimensional analysis framework, combining natural language processing metrics such as information entropy and cross-turn semantic similarity, based on DeBERTa tokenizer \citep{heDeBERTaDecodingenhancedBERT2020} and DeBERTa v3 \citep{heDeBERTaV3ImprovingDeBERTa2021} for analysis. To ensure the robustness of results, we used early BERT models \citep{devlinBERTPretrainingDeep2018} as a control benchmark. Under the MimiTalk Framework, we conducted bilingual corpus comparative analysis of 121 English semi-structured AI interviews (each with transcribed text exceeding 3000 characters) against human-guided interview texts from the MediaSum dataset \citep{zhuMediaSumLargescaleMedia2021}, and tested for significant differences through t-tests. 

\textit{1. Information Entropy}

Information entropy, as a core metric for measuring text complexity and information content, has wide application value in natural language processing and social science fields. Higher entropy reflects greater informational complexity and potentially enhanced interview efficiency \citep{dahlQuantifyingInformationContent2010, friesnerInformationEntropyScale2021}. This approach has proven particularly effective in research contexts where measuring information diversity and content richness is essential for evaluating response quality \citep{homolaMeasureSurveyMode2016, keykhaAdvantagesChallengesElectronic2025}. For example, Dahl and Østerås \citep{dahlQuantifyingInformationContent2010} applied Shannon entropy to survey data, proposing the concept of information efficiency, which is the ratio of empirical entropy to maximum achievable entropy.

We use the DeBERTa base model \citep{heDeBERTaDecodingenhancedBERT2020} for subword tokenization, which provides superior performance in handling mixed-language content compared to traditional word-based approaches.

We calculate the word-level information entropy using the Shannon entropy formula:
\begin{equation}
  H(X) = -\sum_{i=1}^{n} p(x_i) \log_2 p(x_i)
\end{equation}
where $H(X)$ represents the information entropy, $p(x_i)$ is the probability of the $i$-th token, and $\log_2$ denotes the logarithm to base 2.

For each token $i$:
\begin{equation}
  p(x_i) = \frac{\text{count}(x_i)}{\text{total\_tokens}}
\end{equation}

To eliminate the influence of text length on entropy values, we normalize all transcripts to the minimum length across all interviews using truncation. Specifically, if $L_{\min}$ is the minimum token count across all interviews, we truncate longer texts to $L_{\min}$ tokens while preserving the original order of tokens.

The higher resulting entropy values indicate greater lexical diversity and information richness. An entropy of 0 would indicate a text containing only one repeated token, while higher entropy values suggest more diverse vocabulary usage.

To measure the depth and diversity of responses elicited by AI-conducted interviews, we calculate the entropy values separately for interviewer text, interviewee text, and the overall text. Using the same tokenizer, we also calculate sentence length as an auxiliary reference metric. We calculate descriptive statistics (mean, standard deviation, minimum, maximum) for entropy and sentence length across AI-conducted and human-conducted interviews, and assess statistical significance using independent samples t-tests with Welch's correction for unequal variances.

\textit{2. Semantic Similarity}

Semantic similarity measures the degree of conceptual relatedness between textual units, quantifying how closely different utterances align in meaning rather than just lexical overlap. This metric is crucial for evaluating interview quality because it captures conversational coherence—the extent to which participants maintain thematic consistency and engage meaningfully with each other's contributions. High semantic similarity indicates coherent dialogue flow, where questions build upon previous responses and participants stay on-topic, while low similarity may suggest fragmented or disjointed conversations.

In interview contexts, semantic similarity serves three key purposes: measuring how consistently individual speakers maintain thematic focus within their own contributions, indicating depth of engagement; evaluating how well interviewer questions relate to interviewee responses, reflecting conversational quality and mutual understanding; providing a holistic assessment of interview structure and flow. These measures are particularly valuable for comparing AI and human interviewers because they reveal whether AI systems can maintain the natural conversational dynamics that characterize effective human-led interviews.

In social science and human-computer interaction research, the potential of BERT and its derivative models has been widely validated. Fang et al. \citep{fangEvaluatingConstructValidity2022} constructed a validity framework, proving the superiority of BERT embedding technology in representing survey questions, with stronger semantic capture capabilities compared to traditional methods, providing technical assurance for semantic analysis in AI interviews. Grandeit et al. \citep{grandeitUsingBERTQualitative2020} applied BERT to qualitative analysis of psychosocial online counseling, developing an automated coding system with over 50 categories, significantly reducing manual annotation workload. This indicates that AI interviews have high efficiency and scalability in processing large-scale text data. Wankmüller's \citep{wankmullerIntroductionNeuralTransfer2021} research further compared Transformer models with traditional machine learning algorithms, finding that BERT can maintain high prediction accuracy even with limited data, which is particularly crucial for AI interview scenarios. Mostafavi et al. \citep{mostafaviContextualEmbeddingsSociological2025} developed the BERTNN method, extending emotion dictionaries through BERT's contextual embeddings, providing efficient tools for emotion analysis in sociological research and greatly reducing time costs of traditional surveys. Additionally, Aral and Dhillon's \citep{aralWhatExactlyNovelty2023} research explored the sources of information novelty in network structures, pointing out the advantages of weak connections in transmitting unique information, providing theoretical inspiration for understanding AI interviews' role in information transmission. Keykha et al. \citep{keykhaAdvantagesChallengesElectronic2025} analyzed the advantages and disadvantages of electronic examinations through Shannon entropy, revealing the potential and challenges of technology-driven tools in educational assessment, providing reference for automated assessment in AI interviews.

We employ pre-trained transformer models for sentence-level semantic similarity analysis. Specifically, we use the DeBERTa v3 model \citep{heDeBERTaV3ImprovingDeBERTa2021} and BERT base model \citep{devlinBERTPretrainingDeep2018} to generate sentence embeddings, then calculate cosine similarity between sentence pairs.

For sentence segmentation, we split text using punctuation markers ($[.!?]+$) and speaker indicators. Each sentence $s_i$ is tokenized and embedded using the transformer model:
\begin{equation}
\mathbf{e}_i = \text{Model}(s_i) \in \mathbb{R}^{768}
\end{equation}

where $\mathbf{e}_i$ is the 768-dimensional sentence embedding obtained by mean pooling over the last hidden states of the transformer model.

In the semantic similarity analysis, we separately calculated the cosine similarity of sentences within the interviewer, within the interviewee, and between the two speakers. Specifically, let $\mathbf{e}_i^{\text{int}}$ denote the embedding vector of the $i$-th interviewer sentence, $N_{\text{int}}$ the number of interviewer sentences, $\mathbf{e}_j^{\text{intv}}$ the embedding vector of the $j$-th interviewee sentence, and $N_{\text{intv}}$ the number of interviewee sentences. The formulas for the three types of similarity are as follows:
\begin{equation}
\text{Sim}_{\text{int}} = \frac{1}{N_{\text{int}}(N_{\text{int}}-1)/2} \sum_{i<j} \frac{\mathbf{e}_i^{\text{int}} \cdot \mathbf{e}_j^{\text{int}}}{\|\mathbf{e}_i^{\text{int}}\| \|\mathbf{e}_j^{\text{int}}\|}
\end{equation}
\begin{equation}
\text{Sim}_{\text{intv}} = \frac{1}{N_{\text{intv}}(N_{\text{intv}}-1)/2} \sum_{i<j} \frac{\mathbf{e}_i^{\text{intv}} \cdot \mathbf{e}_j^{\text{intv}}}{\|\mathbf{e}_i^{\text{intv}}\| \|\mathbf{e}_j^{\text{intv}}\|}
\end{equation}
\begin{equation}
\text{Sim}_{\text{cross}} = \frac{1}{N_{\text{int}} N_{\text{intv}}} \sum_{i,j} \frac{\mathbf{e}_i^{\text{int}} \cdot \mathbf{e}_j^{\text{intv}}}{\|\mathbf{e}_i^{\text{int}}\| \|\mathbf{e}_j^{\text{intv}}\|}
\end{equation}

Here, $\cdot$ denotes the dot product, and $\|\cdot\|$ denotes the L2 norm.

The cosine similarity ranges from -1 to 1, where 1 indicates identical semantic content, 0 represents orthogonal (unrelated) content, and -1 denotes opposite semantic content.

For robustness testing, we compare results across multiple transformer models (BERT and DeBERTa v3) to ensure consistency of findings.

We calculate descriptive statistics (mean, standard deviation, minimum, maximum, median) for each similarity type across AI-conducted and human-conducted interviews, and assess statistical significance using independent samples t-tests with Welch's correction for unequal variances.

\textit{3. Causal identification}

We employ Propensity Score Matching (PSM) to identify the causal effects of AI-conducted interviews on linguistic characteristics. This quasi-experimental approach addresses potential selection bias by balancing observable confounders between AI and human interview groups.

We construct covariates using interview-level characteristics that affect interview type selection but are not directly influenced by AI technology:
\begin{equation}
X_i = [\text{total\_tokens}_i, \text{total\_sentences}_i, \text{PC1}_i, \text{PC2}_i, \text{PC3}_i]
\end{equation}

where $\text{total\_tokens}_i$ and $\text{total\_sentences}_i$ represent interview length and dialogue turns, and $\text{PC1}_i, \text{PC2}_i, \text{PC3}_i$ are principal components from topic embedding analysis. We apply log transformation, polynomial features, and quantile transformation to improve covariate distributions.

We estimate propensity scores using logistic regression:
\begin{equation}
P(T_i = 1 | X_i) = \frac{\exp(\beta_0 + \beta_1 X_{i1} + ... + \beta_k X_{ik})}{1 + \exp(\beta_0 + \beta_1 X_{i1} + ... + \beta_k X_{ik})}
\end{equation}

where $T_i = 1$ indicates AI-conducted interview and $T_i = 0$ indicates human-conducted interview.

We employ Kernel Matching with Gaussian kernel function and adaptive caliper $c = 0.2 \times \text{SD}(PS)$:
\begin{equation}
K(d_{ij}) = \exp\left(-\frac{d_{ij}^2}{2h^2}\right)
\end{equation}

where $d_{ij} = |PS_i - PS_j|$ is the propensity score distance and $h = 0.1$ is the bandwidth parameter.

On the matched sample, we estimate Average Treatment Effect (ATE) using Ordinary Least Squares:
\begin{equation}
Y_i = \alpha + \beta T_i + \gamma_1 X_{i1} + ... + \gamma_k X_{ik} + \epsilon_i
\end{equation}

where $Y_i$ is the outcome variable (entropy, sentence length, or semantic similarity), $T_i$ is the treatment indicator, and $\beta$ is the ATE estimate.

We conduct placebo tests using digit ratio and sentence count variables theoretically unaffected by AI interview technology, and perform alternative metric testing using DeBERTa v3 embeddings for semantic similarity. Our causal identification relies on unconfoundedness ($Y(1), Y(0) \perp T | X$), overlap ($0 < P(T = 1 | X) < 1$), and SUTVA (no interference between units). We assess covariate balance using Standardized Mean Difference (SMD) and report ATE estimates with standard errors and 95\% confidence intervals.

We address potential endogeneity issues through comprehensive length normalization mechanisms. For semantic similarity analysis, we implement sentence-level normalization, embedding dimension normalization, and similarity calculation normalization, ensuring that \texttt{total\_tokens} and \texttt{total\_sentences} as covariates do not introduce endogeneity problems. For information entropy analysis, we apply text length normalization, text truncation standardization, and information entropy calculation normalization, where entropy values are computed based on standardized text lengths and all interview texts are truncated to the same minimum length. These normalization procedures eliminate the potential endogeneity concerns that could arise from text length variations affecting our outcome variables.

\subsection{Study 3: Comparative Analysis of Human-led vs. AI-conducted Interviews}

To further validate the deep insight capabilities of AI systems, we evaluated whether AI systems can achieve or even surpass the depth and authenticity of traditional human interviewers by comparing human-led and AI-led interview approaches. We chose a more precise controlled variable method, with this design philosophy based on a comparative case study methodological framework that has been proven effective in evaluating different data analysis methods \citep{akmanHumanResearcherVs2025}. We therefore conducted human interviews and AI semi-structured interviews under the MimiTalk framework using consistent outlines, followed by blind thematic/content analysis of verbatim transcripts.

We recruited 10 academic researchers across diverse disciplinary backgrounds, ensuring methodological rigor through representational diversity. Participants included PhD students, lecturers, postdoctoral researchers, and professors with 0.5 to 10 years of research experience, all bilingual in Chinese and English. Disciplinary backgrounds spanned philosophy (including Marxist theory), microbiology, computer science (vehicular networking and model optimization), economics (economic history), strategic management (innovation and human-AI collaboration), atmospheric science, speech recognition, and image processing. Detailed demographic and research background information is provided in Appendix B.

Semi-structured interviews, as an important method in qualitative research, have wide application in social science research. This method combines the standardization advantages of structured interviews with the flexibility of open interviews, allowing researchers to conduct in-depth follow-up questions based on respondents' answers while maintaining focus on core themes \citep{brinkmannQualitativeInterviewing2013}. In AI interview systems, semi-structured design is particularly important because it provides sufficient guidance framework for AI systems while retaining the ability to adjust questions based on conversational dynamics. We conducted blind thematic/content analysis of verbatim transcripts, with reviewers (unaware of interview sources) using reflective thematic analysis methods to evaluate the thematic depth, emotional subtlety, and information richness of texts. This blind review method has been proven to effectively reduce subjective bias and is widely used in medical and social science qualitative research. For example, Sakaguchi et al.'s \citep{sakaguchiEvaluatingChatGPTQualitative2025} study comparing ChatGPT with humans in Japanese clinical contexts showed that AI achieved over 80\% agreement on descriptive themes but only about 30\% on cultural emotional themes, inspiring us to pay special attention to AI's boundaries in handling sensitive topics. Similarly, Prescott et al. \citep{prescottComparingEfficacyEfficiency2024} compared humans and generative AI in qualitative thematic analysis, finding that AI achieved 71\% thematic consistency in inductive analysis and significantly reduced analysis time (20 minutes vs. 567 minutes), providing empirical foundation for emphasizing the necessity of hybrid modes. Gibson and Beattie \citep{gibsonMoreLessHuman2024} further pointed out that AI struggles to replicate the richness of human experience in online qualitative research, emphasizing the role of theoretical frameworks in identifying AI-generated data, which aligns with our blind review method.

In interviews conducted under the Mimitalk Framework, we introduced more accurate outlines and protocols, informing the AI system in advance about the content researchers want to explore and potential issues they might encounter through examples and guiding words, combined with the AI system's self-adjustment capabilities, making interviews more natural and in-depth. This prompt engineering method has been proven to significantly improve analysis efficiency and accuracy in AI-assisted qualitative analysis \citep{akmanHumanResearcherVs2025}. Our prompts designed four themes, including: AI ethics and responsibility, AI potential risks and challenges, AI practical applications and effects, and AI future development and prospects.

Interdisciplinary research design has significant importance in evaluating the effectiveness of AI systems in different field applications. By recruiting researchers from different disciplinary backgrounds, we were able to assess the adaptability and effectiveness of AI interview systems in diverse research contexts. This design method not only enhanced the external validity of research results but also provided a comprehensive perspective for understanding AI system performance in different professional fields \citep{stokolsScienceTransdisciplinaryAction2006}. The participation of bilingual researchers further ensured the applicability of research results in cross-cultural contexts, which is crucial for evaluating the potential of AI systems in globalized research environments.

English Prompts (translated): {Brief understanding: research field, years of work experience, working language
1. If you were to rate the academic research value of AI in your field on a scale of 1-5 (1=completely useless, 5=transformative), what would your score be? Could you briefly explain the reasons? Please explain from perspectives such as research efficiency, innovation, and quality of outcomes.
2. Compared to traditional methods, has AI assistance changed your time allocation patterns? In which specific work stages or steps have you felt changes in time allocation: for example, in literature search, data analysis, writing drafts, etc., have there been obvious time reallocations?
3. From 0\% (completely distrust) to 100\% (completely trust), what is your reliability rating for AI-generated academic content? Please explain this value with specific scenarios [If below 80\%, ask: How would you verify AI-generated content when reliability is insufficient?]
4. In literature review/experimental design/data collection/paper writing/outcome review, which stage do you think AI is most suitable to intervene in? Which is least suitable? Please provide examples. Are there any unlisted stages where you think AI might be applicable?
5. What are your expectations and concerns about the future of AI in academic research? What opportunities might it bring? What risks should we be vigilant about? (Please provide examples)
6. In your research process, have AI tools ever had a significant impact on your conclusions or research direction? Please provide examples. Have you ever adjusted research methods, modified hypotheses, or discovered new research questions because of AI suggestions?}

\section{Results}

\subsection{Study 2}
\subsubsection{Descriptive Statistics}
Our comprehensive analysis of 121 AI interviews compared to 1,271 human interviews from the MediaSum dataset reveals significant differences across three key dimensions: information entropy, token usage patterns, and semantic similarity.

\textbf{Information Entropy Analysis}

\begin{figure}[h]
  \centering
  \begin{subfigure}[b]{0.32\textwidth}
    \centering
    \includegraphics[width=\textwidth]{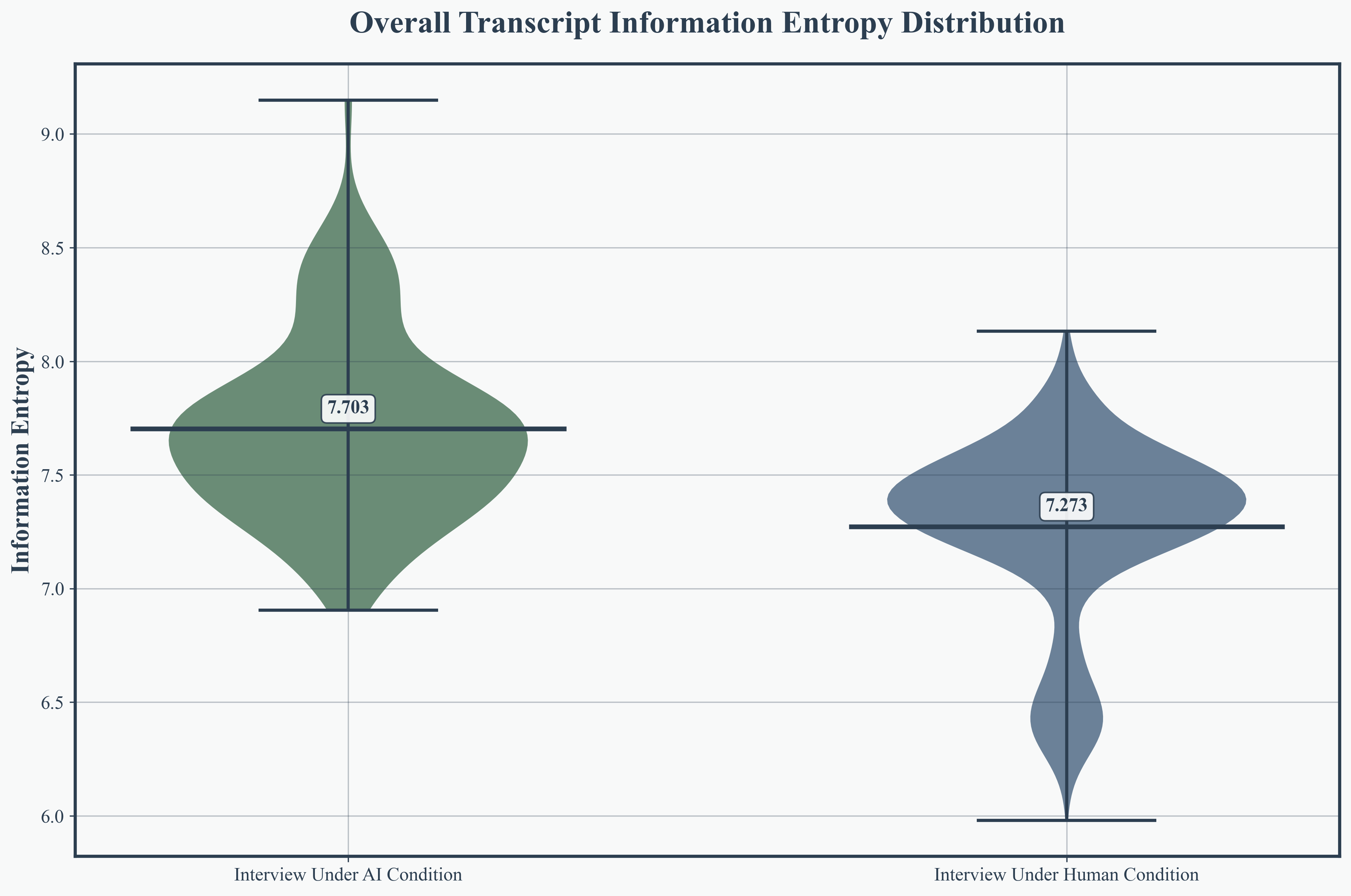}
    \caption{Overall transcript entropy}
    \label{fig:entropy_overall}
  \end{subfigure}
  \hfill
  \begin{subfigure}[b]{0.32\textwidth}
    \centering
    \includegraphics[width=\textwidth]{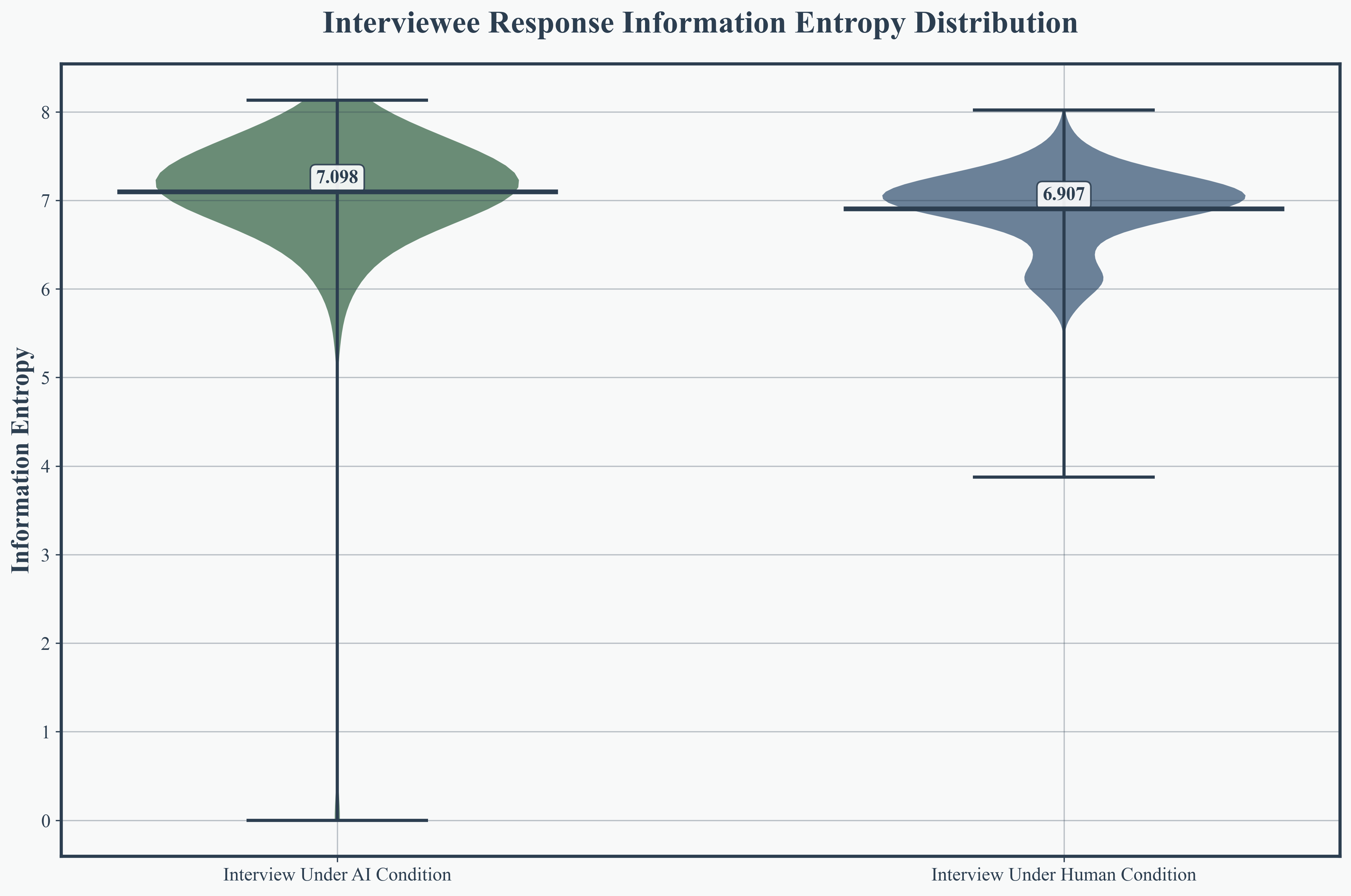}
    \caption{Interviewee response entropy}
    \label{fig:entropy_interviewee}
  \end{subfigure}
  \hfill
  \begin{subfigure}[b]{0.32\textwidth}
    \centering
    \includegraphics[width=\textwidth]{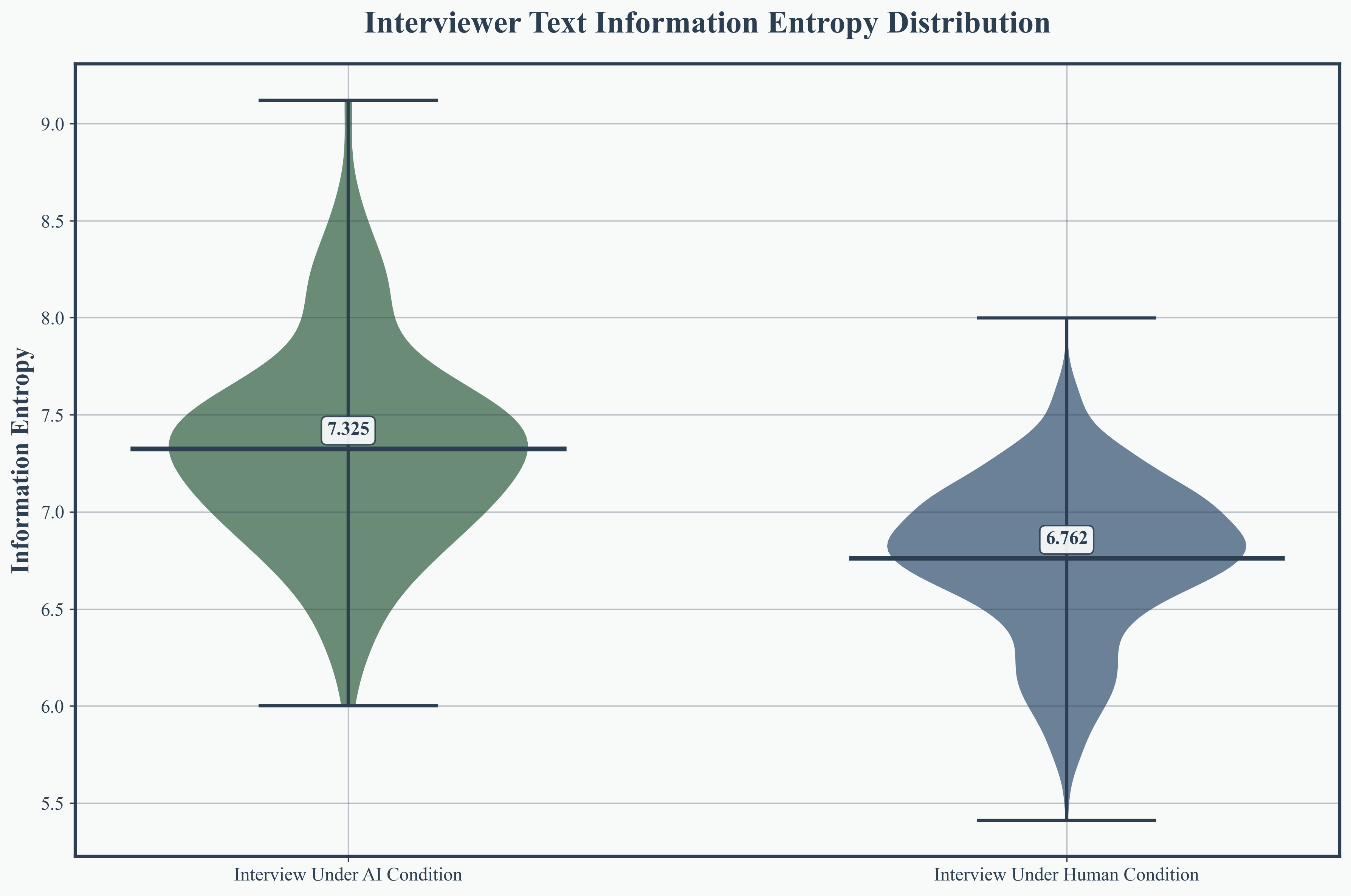}
    \caption{Interviewer text entropy}
    \label{fig:entropy_interviewer}
  \end{subfigure}
  \caption{Information entropy distributions comparing AI and human interviews. (a) Overall transcript entropy: AI interviews demonstrate significantly higher linguistic diversity (7.703 ± 0.399) compared to human interviews (7.273 ± 0.395), representing a 5.9\% increase in vocabulary richness. (b) Interviewee response entropy: AI interviewees exhibit higher entropy (7.098 ± 0.791) than human interviewees (6.907 ± 0.446), indicating more diverse vocabulary usage in responses. (c) Interviewer text entropy: AI interviewers show substantially higher entropy (7.325 ± 0.512) compared to human interviewers (6.762 ± 0.401), representing an 8.3\% increase in question diversity.}
\end{figure}

\begin{figure}[h]
  \centering
  \begin{subfigure}[b]{0.32\textwidth}
    \centering
    \includegraphics[width=\textwidth]{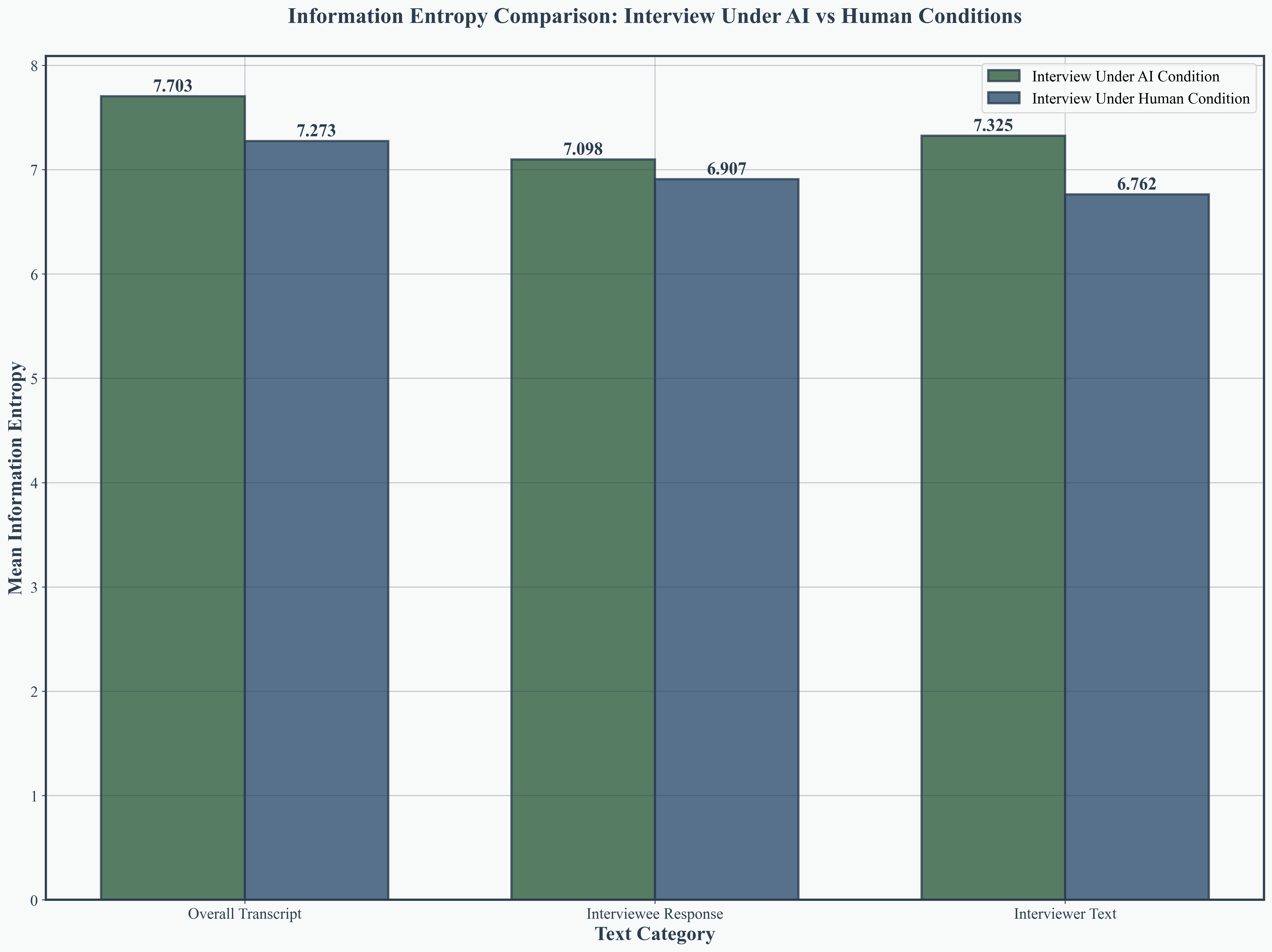}
    \caption{Comprehensive entropy comparison}
    \label{fig:comprehensive_entropy}
  \end{subfigure}
  \hfill
  \begin{subfigure}[b]{0.32\textwidth}
    \centering
    \includegraphics[width=\textwidth]{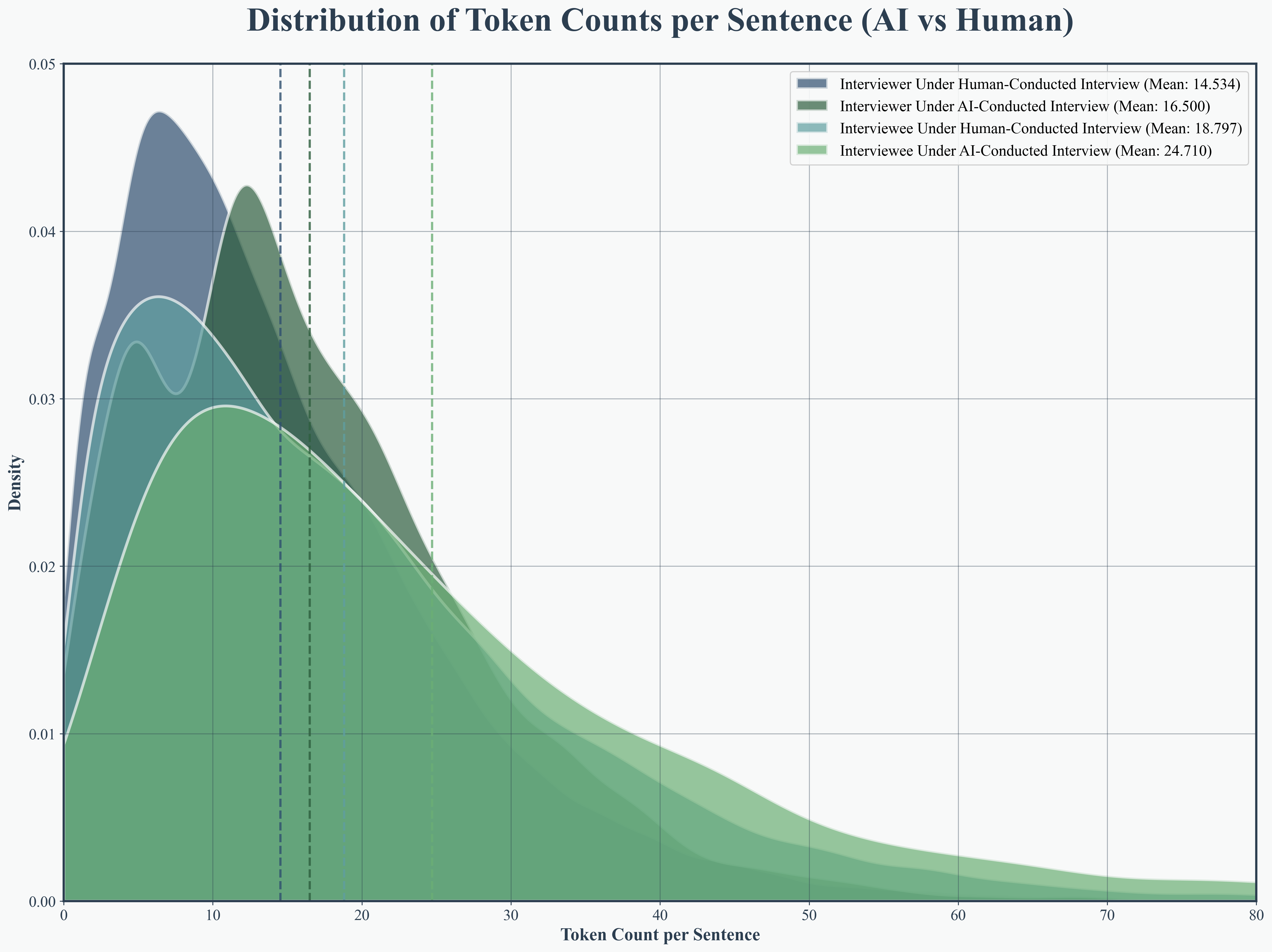}
    \caption{Token count KDE}
    \label{fig:token_kde}
  \end{subfigure}
  \hfill
  \begin{subfigure}[b]{0.32\textwidth}
    \centering
    \includegraphics[width=\textwidth]{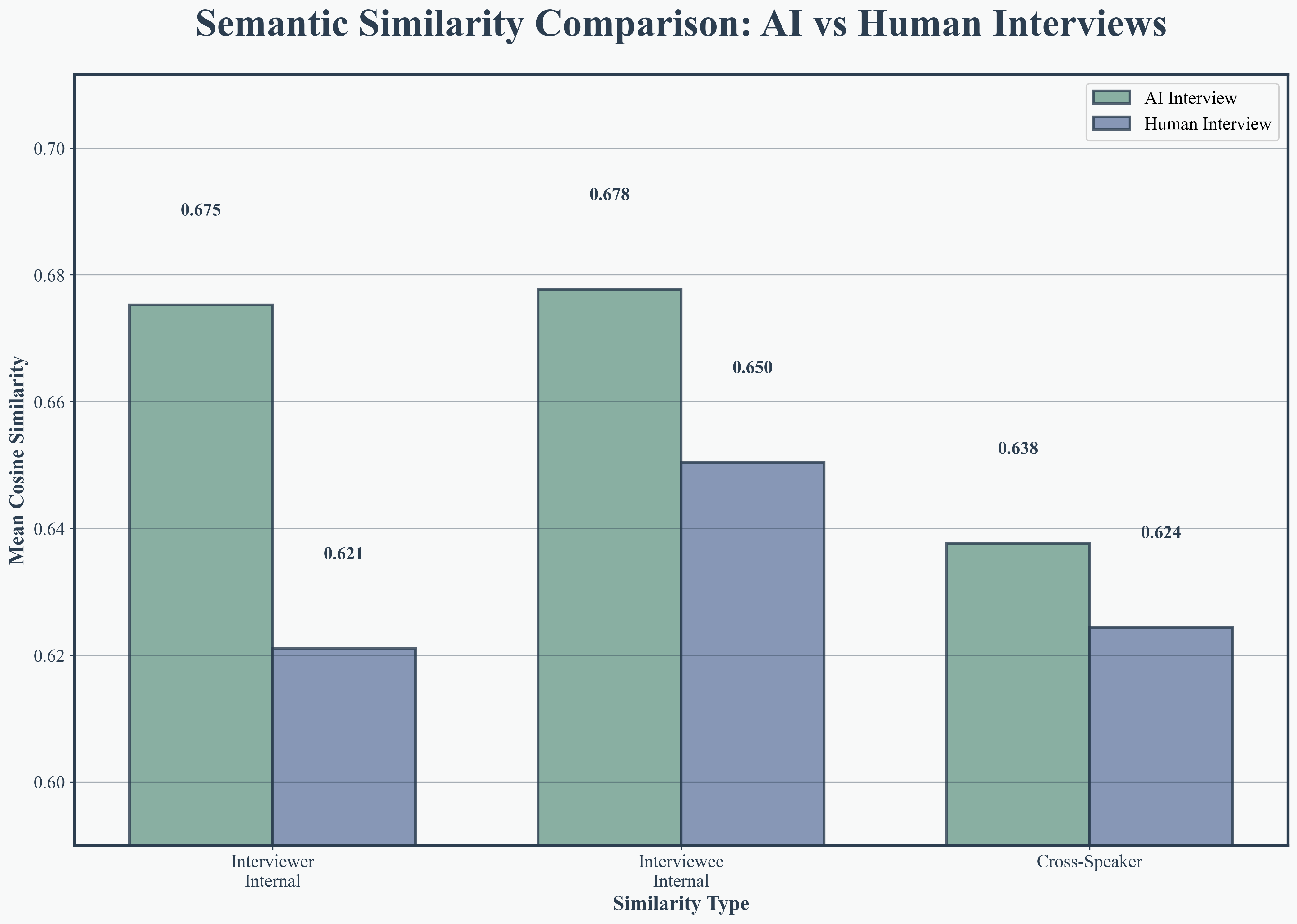}
    \caption{Similarity comparison}
    \label{fig:similarity_comparison}
  \end{subfigure}
  \caption{Comprehensive analysis summaries. (a) Comprehensive information entropy comparison across all categories: The violin plots demonstrate that AI interviews consistently achieve higher linguistic diversity across overall transcripts, interviewer questions, and interviewee responses, with tighter distributions indicating more reliable performance. (b) Kernel density estimation of token count distributions: The density plots reveal distinct patterns where AI interviews show broader, more right-skewed distributions for both interviewers and interviewees, indicating greater variability in response lengths compared to human interviews. (c) Comprehensive semantic similarity comparison across all categories: The violin plots demonstrate that AI interviews consistently outperform human interviews in semantic coherence across interviewer internal, interviewee internal, and cross-speaker similarity measures.}
\end{figure}

Information entropy analysis demonstrates that AI-conducted interviews consistently exhibit higher linguistic diversity across all measured categories. Overall transcript entropy shows AI interviews achieving significantly higher values (7.703 ± 0.399) compared to human interviews (7.273 ± 0.395), representing a 5.9\% increase in vocabulary diversity (Figure \ref{fig:entropy_overall}). This pattern extends to both speaker roles: AI interviewee responses demonstrate higher entropy (7.098 ± 0.791) than human counterparts (6.907 ± 0.446), while AI interviewer text shows the most pronounced difference (7.325 ± 0.512 vs. 6.762 ± 0.401), representing an 8.3\% increase in question diversity (Figures \ref{fig:entropy_interviewee} and \ref{fig:entropy_interviewer}). The comprehensive comparison across all categories confirms AI interviews' superior linguistic richness (Figure \ref{fig:comprehensive_entropy}).

\textbf{Token Usage Patterns}

\begin{figure}[h]
  \centering
  \begin{subfigure}[b]{0.32\textwidth}
    \centering
    \includegraphics[width=\textwidth]{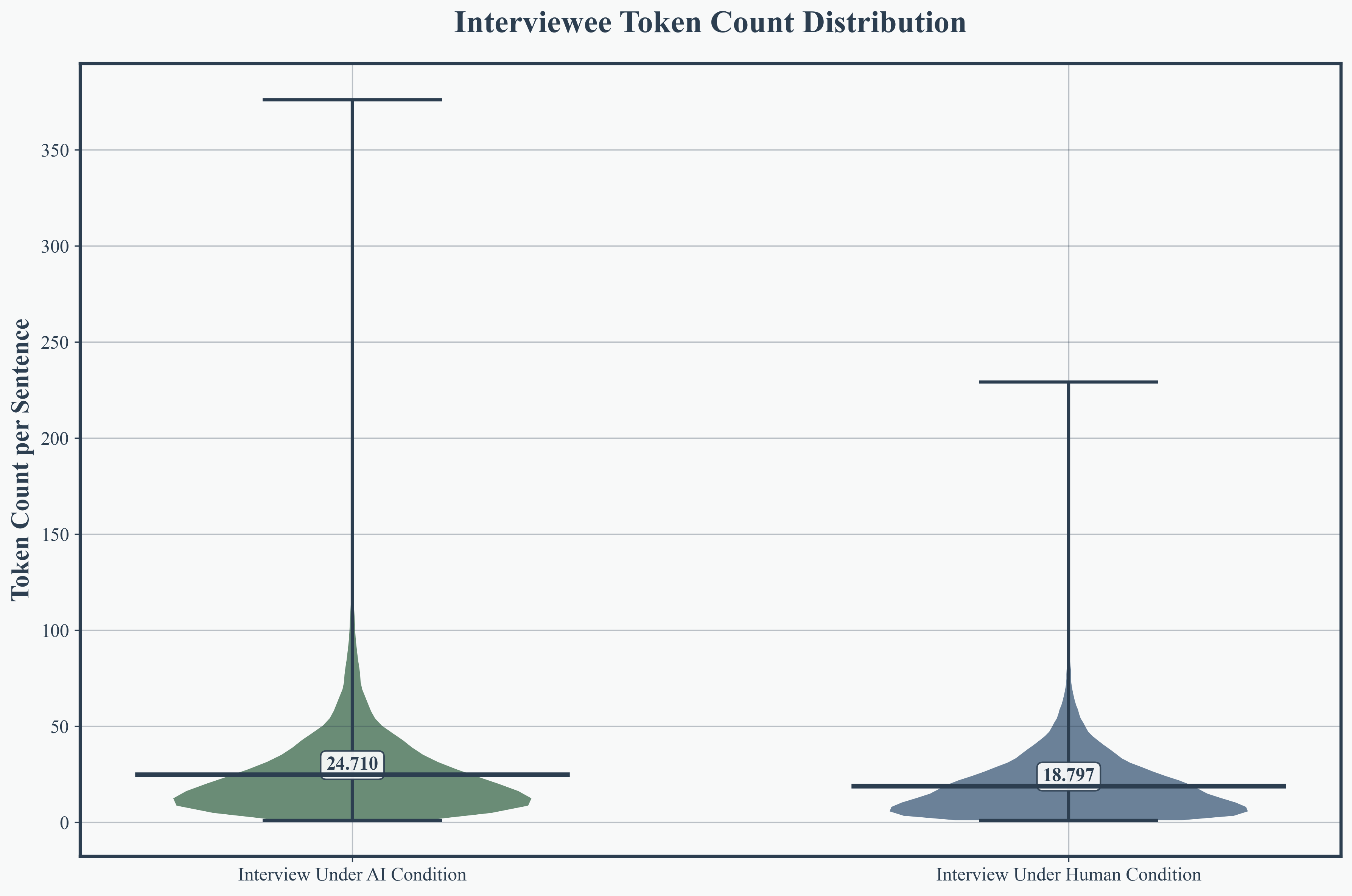}
    \caption{Interviewee response token count}
    \label{fig:token_interviewee}
  \end{subfigure}
  \hfill
  \begin{subfigure}[b]{0.32\textwidth}
    \centering
    \includegraphics[width=\textwidth]{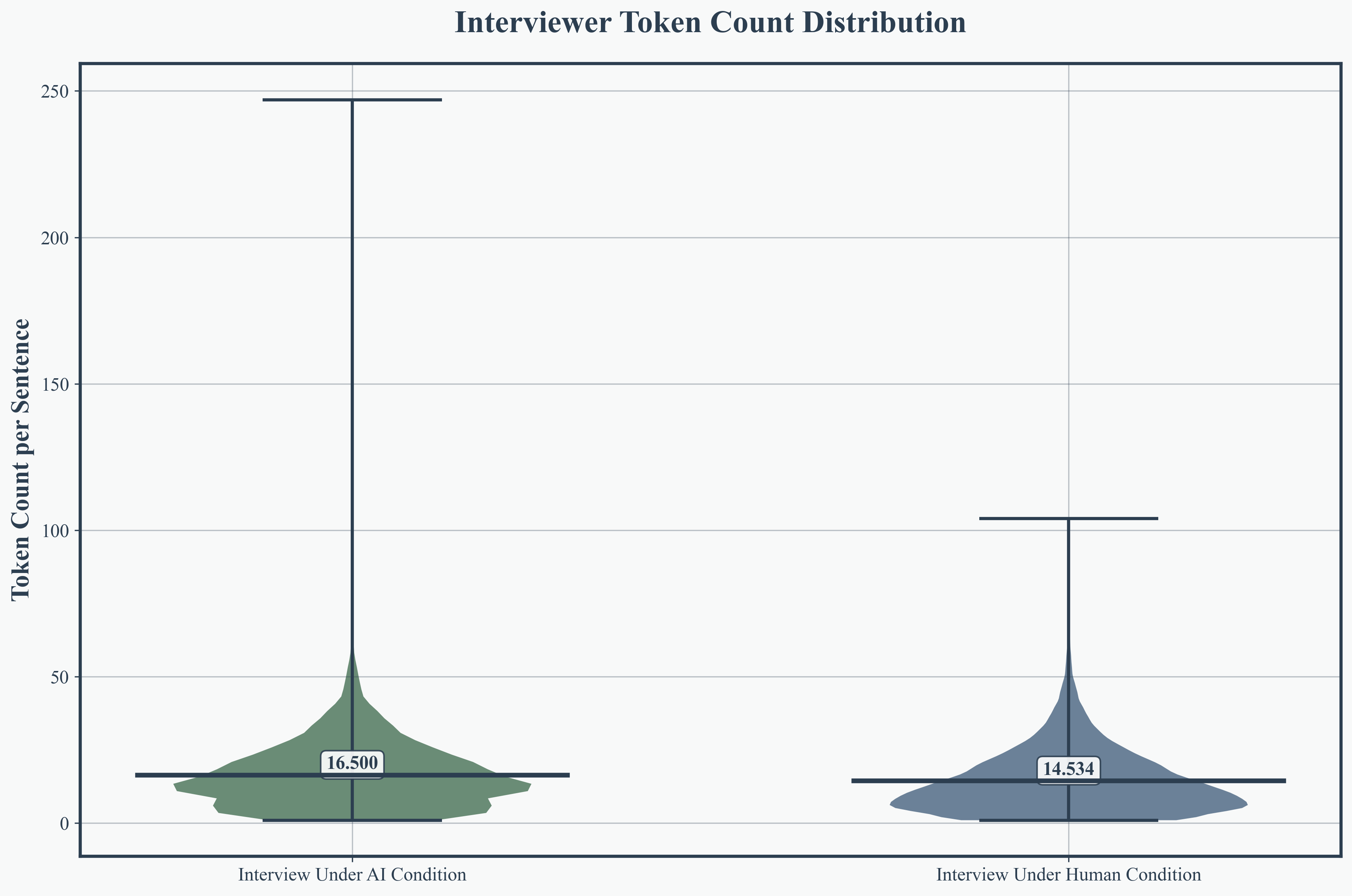}
    \caption{Interviewer question token count}
    \label{fig:token_interviewer}
  \end{subfigure}
  \hfill
  \begin{subfigure}[b]{0.32\textwidth}
    \centering
    \includegraphics[width=\textwidth]{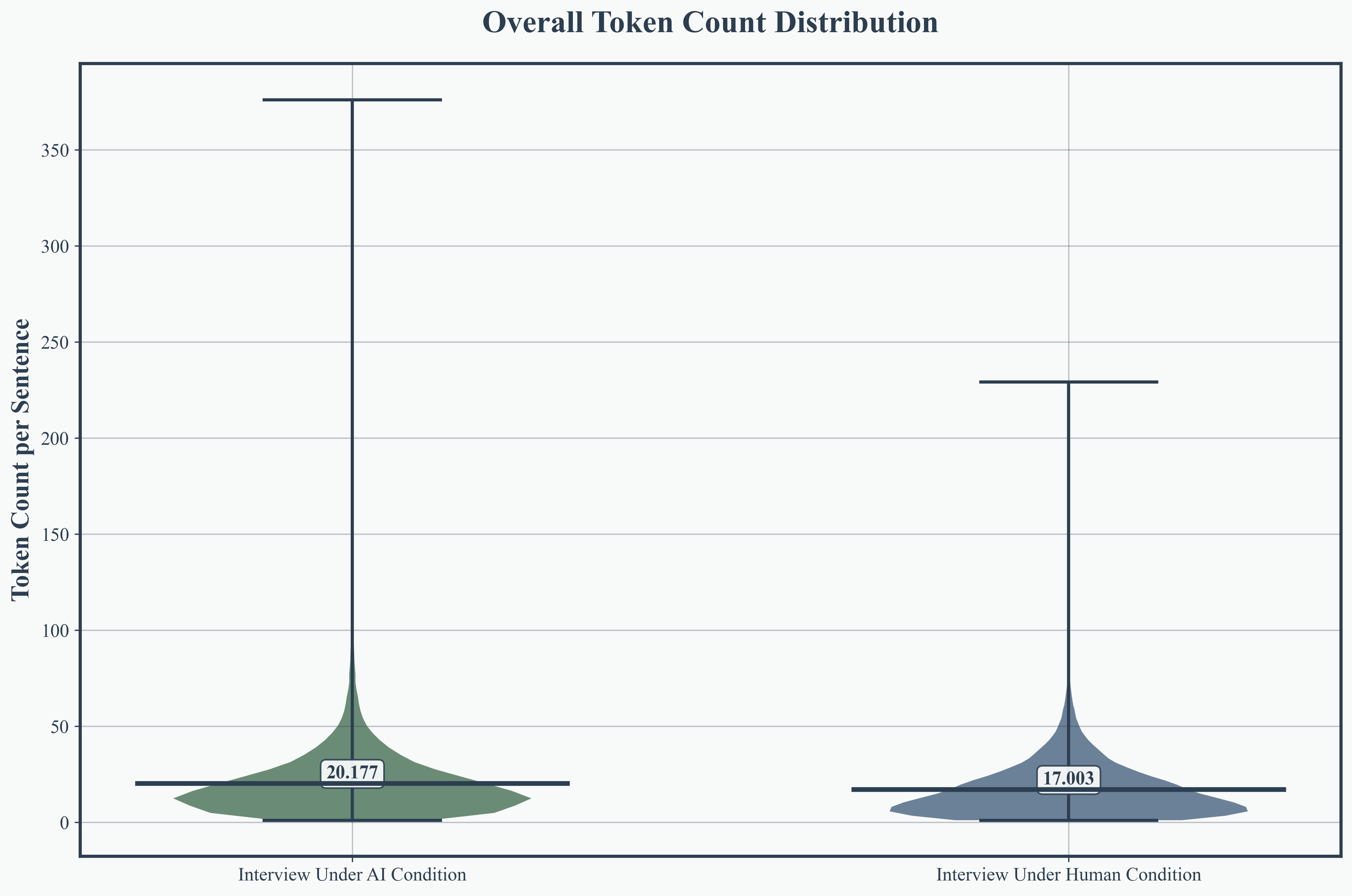}
    \caption{Overall token count distribution}
    \label{fig:token_overall}
  \end{subfigure}
  \caption{Token count distributions in AI and human interviews. (a) Interviewee response token count: AI interviewees produce significantly longer responses (24.710 ± 22.232 tokens per sentence) compared to human interviewees (18.797 ± 15.453 tokens), representing a 31.4\% increase in response length. (b) Interviewer question token count: AI interviewers generate slightly longer questions (16.500 ± 11.614 tokens) than human interviewers (14.534 ± 11.269 tokens), showing a 13.8\% increase. (c) Overall token count distribution: AI interviews demonstrate higher average token counts (20.177 ± 17.678) compared to human interviews (17.003 ± 14.006), with greater variability indicating more diverse response strategies.}
\end{figure}

Token count analysis reveals distinct patterns in response length and variability between AI and human interviews. AI interviewees produce significantly longer responses (24.7 ± 22.2 tokens per sentence) compared to human interviewees (18.8 ± 15.5 tokens), representing a 31.4\% increase in response length (Figure \ref{fig:token_interviewee}). AI interviewers also generate slightly longer questions (16.5 ± 11.6 tokens) than human interviewers (14.5 ± 11.3 tokens), showing a 13.8\% increase (Figure \ref{fig:token_interviewer}). Notably, AI interviews exhibit greater variability in response lengths, particularly among interviewees (standard deviation: 22.2 vs. 15.5 tokens), suggesting more diverse response strategies (Figures \ref{fig:token_overall} and \ref{fig:token_kde}).

\textbf{Semantic Similarity Analysis}

\begin{figure}[h]
  \centering
  \begin{subfigure}[b]{0.32\textwidth}
    \centering
    \includegraphics[width=\textwidth]{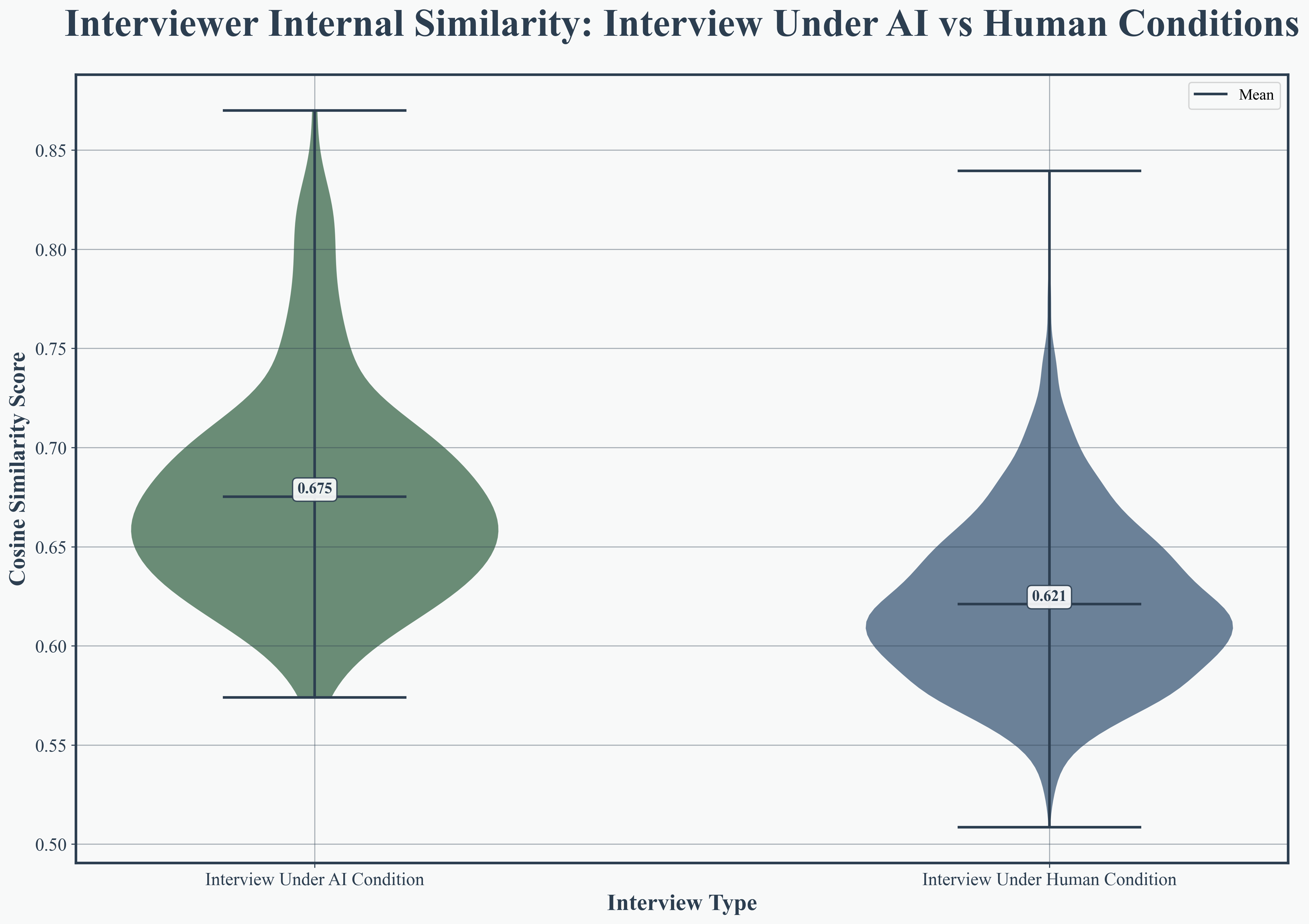}
    \caption{Interviewer internal similarity}
    \label{fig:interviewer_similarity}
  \end{subfigure}
  \hfill
  \begin{subfigure}[b]{0.32\textwidth}
    \centering
    \includegraphics[width=\textwidth]{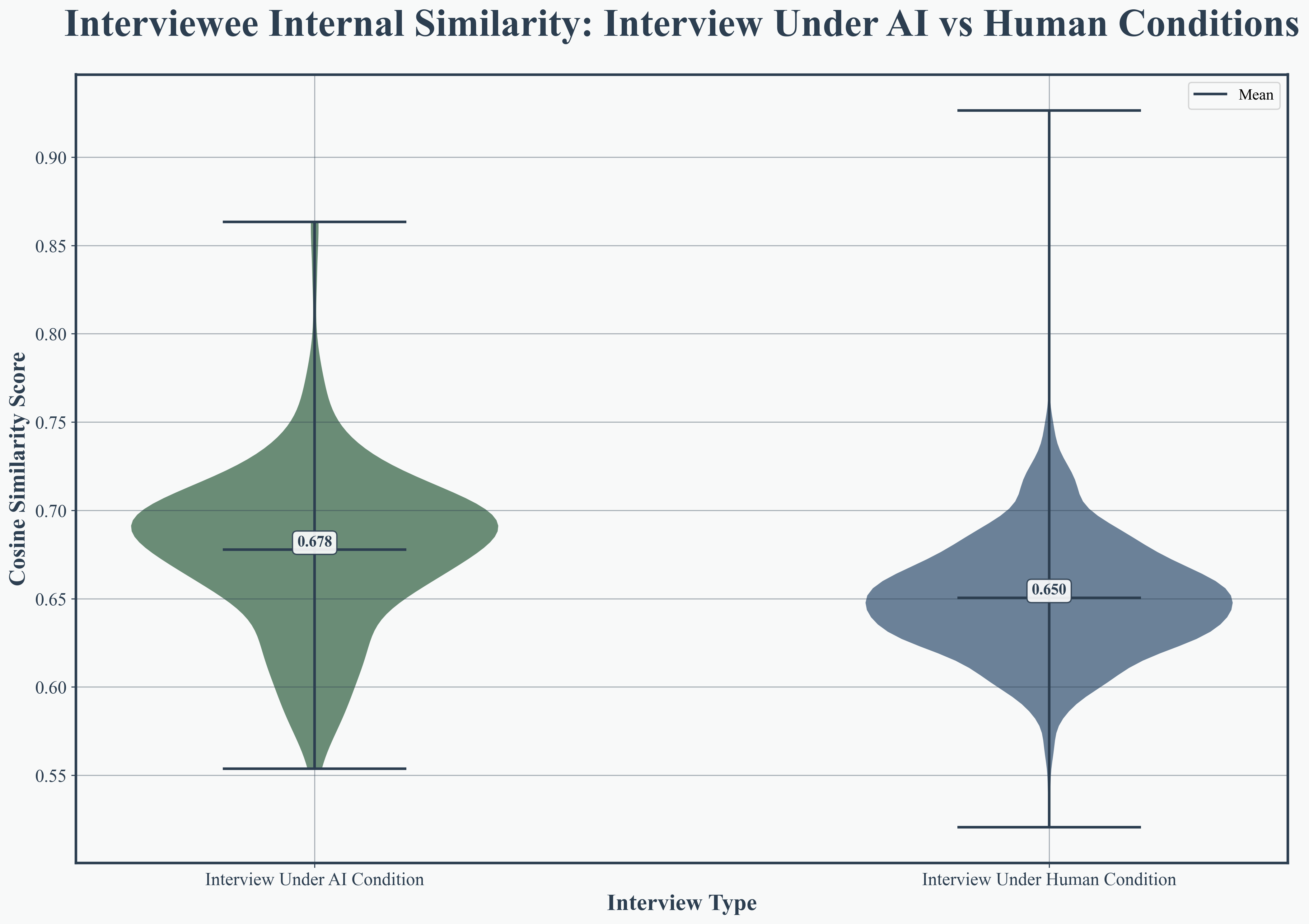}
    \caption{Interviewee internal similarity}
    \label{fig:interviewee_similarity}
  \end{subfigure}
  \hfill
  \begin{subfigure}[b]{0.32\textwidth}
    \centering
    \includegraphics[width=\textwidth]{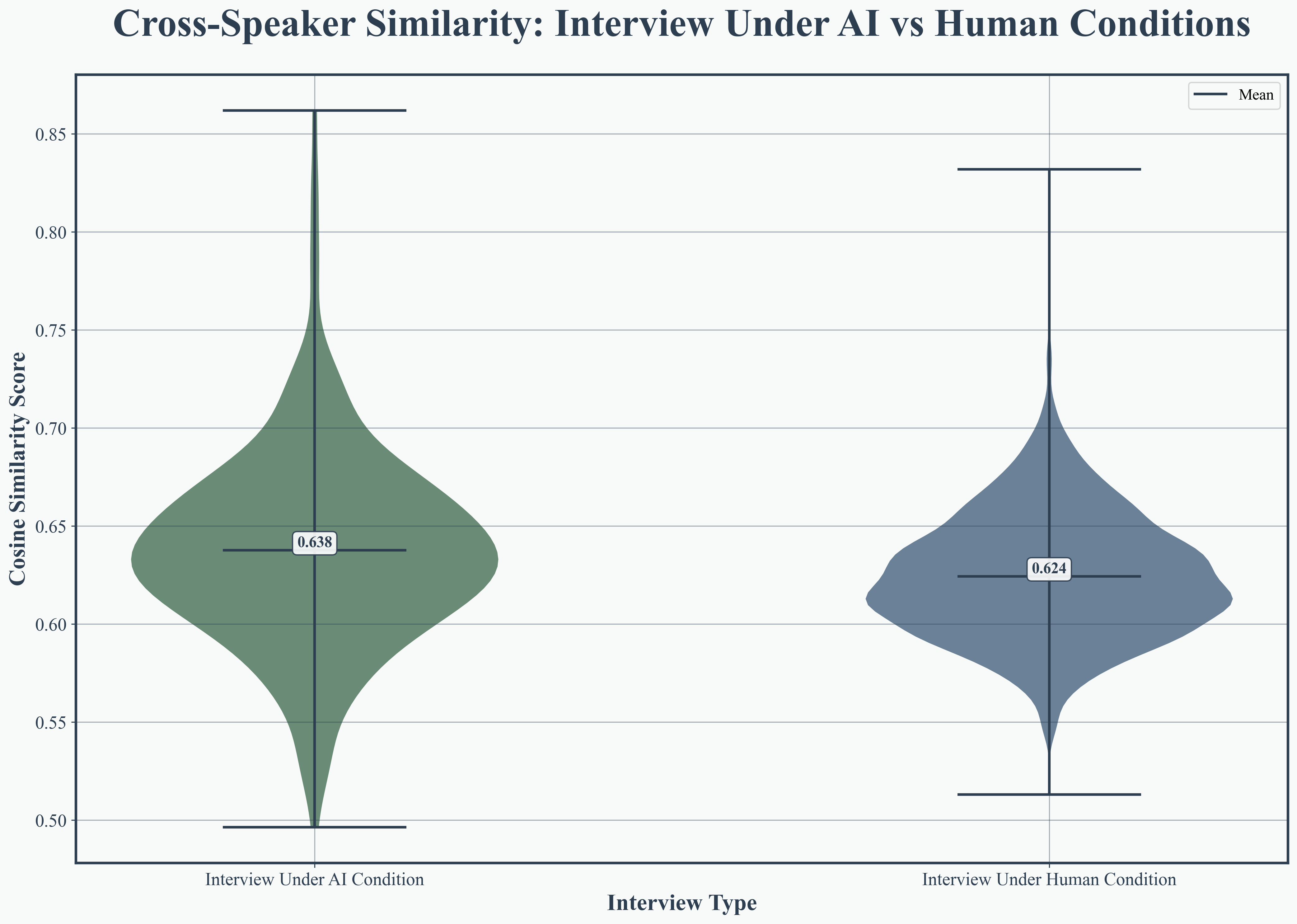}
    \caption{Cross-speaker similarity}
    \label{fig:cross_similarity}
  \end{subfigure}
  \caption{Semantic similarity comparison between AI and human interviews. (a) Interviewer internal similarity: AI interviewers exhibit higher semantic consistency (0.6540 ± 0.0493) than human interviewers (0.5964 ± 0.0265), indicating more coherent questioning strategies by AI. (b) Interviewee internal similarity: AI interviewees show greater internal semantic consistency (0.6860 ± 0.0454) compared to human interviewees (0.6446 ± 0.0278), suggesting more consistent responses within the same interview. (c) Cross-speaker similarity: The semantic alignment between interviewer and interviewee is higher in AI interviews (0.6234 ± 0.0521 vs. 0.6057 ± 0.0218), reflecting an advantage of AI interviews in conversational coherence and topic consistency.}
\end{figure}

Semantic similarity analysis using DeBERTa-v3 embeddings reveals that AI interviews demonstrate higher semantic coherence across all measured dimensions. AI interviewers show higher internal similarity (0.886 ± 0.025) compared to human interviewers (0.814 ± 0.064), indicating more consistent questioning strategies (Figure \ref{fig:interviewer_similarity}). AI interviewees similarly exhibit higher internal similarity (0.892 ± 0.034) than human interviewees (0.847 ± 0.056), suggesting more coherent response patterns within individual interviews (Figure \ref{fig:interviewee_similarity}). Cross-speaker similarity analysis shows AI interviews achieving higher semantic alignment between interviewer and interviewee (0.872 ± 0.029 vs. 0.824 ± 0.048), indicating better conversational coherence (Figure \ref{fig:cross_similarity}). The comprehensive similarity comparison confirms AI interviews' superior semantic consistency across all categories (Figure \ref{fig:similarity_comparison}).

\begin{figure}[h]
  \centering
  \begin{subfigure}[b]{0.32\textwidth}
    \centering
    \includegraphics[width=\textwidth]{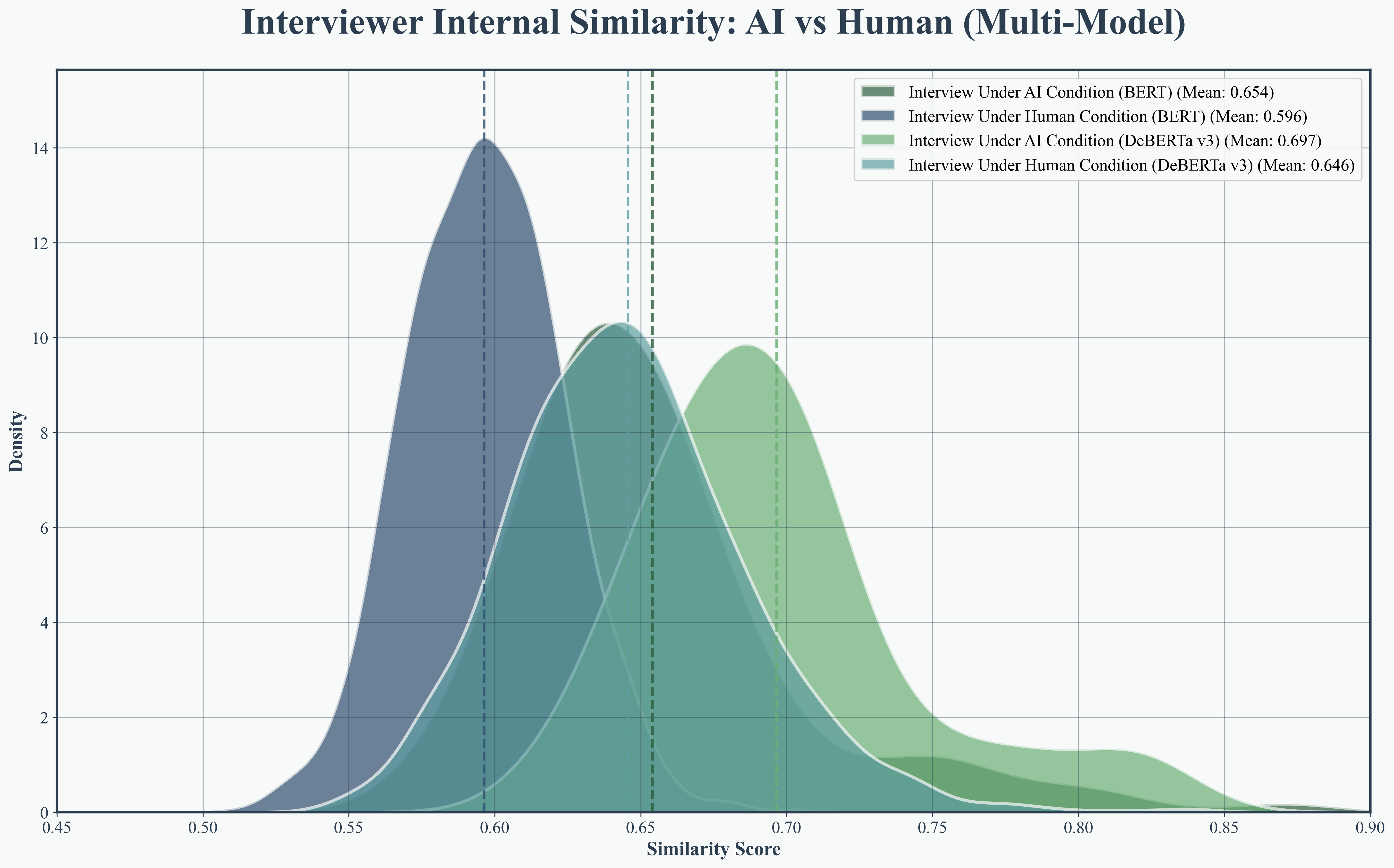}
    \caption{Interviewer internal similarity KDE}
    \label{fig:interviewer_kde}
  \end{subfigure}
  \hfill
  \begin{subfigure}[b]{0.32\textwidth}
    \centering
    \includegraphics[width=\textwidth]{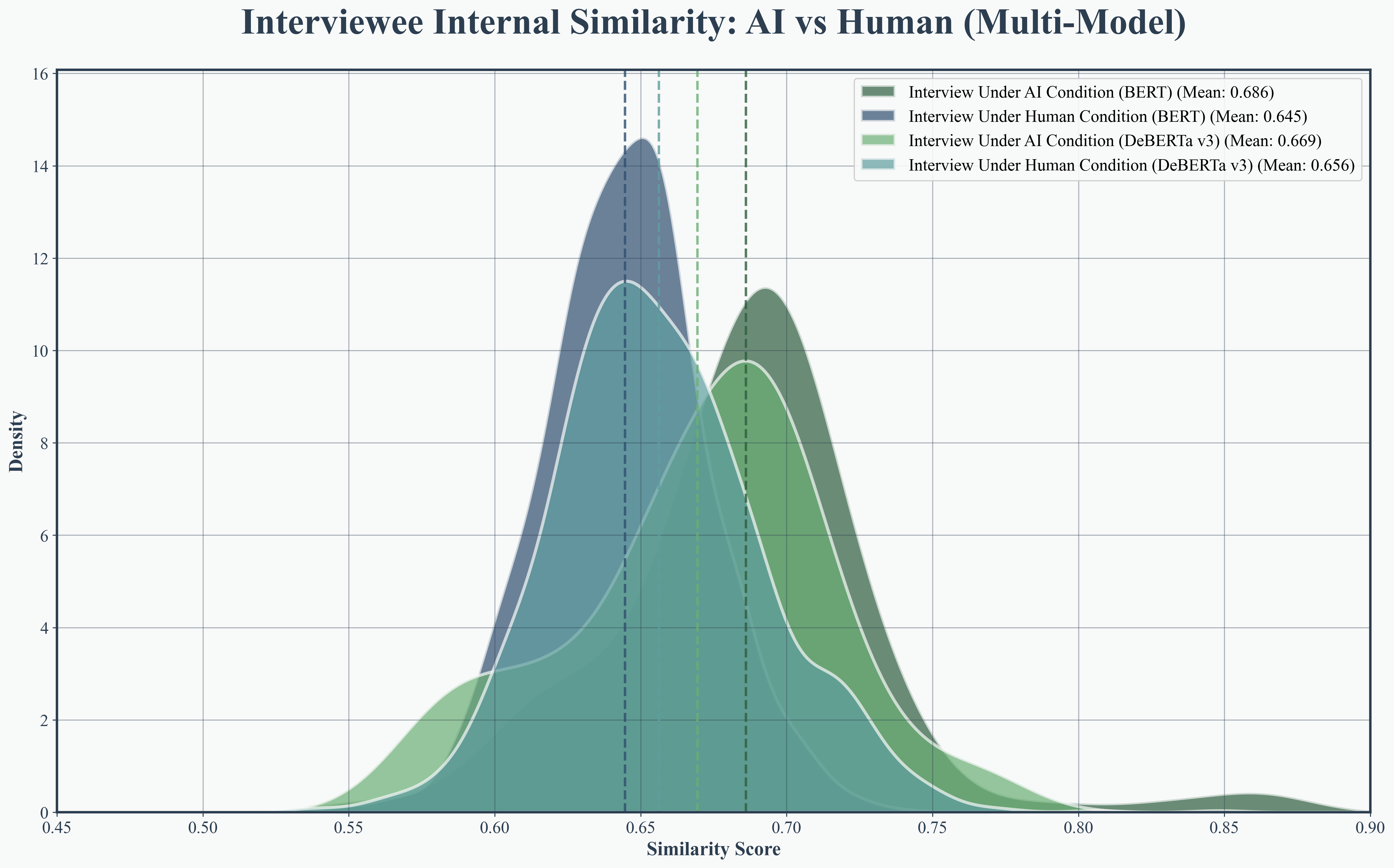}
    \caption{Interviewee internal similarity KDE}
    \label{fig:interviewee_kde}
  \end{subfigure}
  \hfill
  \begin{subfigure}[b]{0.32\textwidth}
    \centering
    \includegraphics[width=\textwidth]{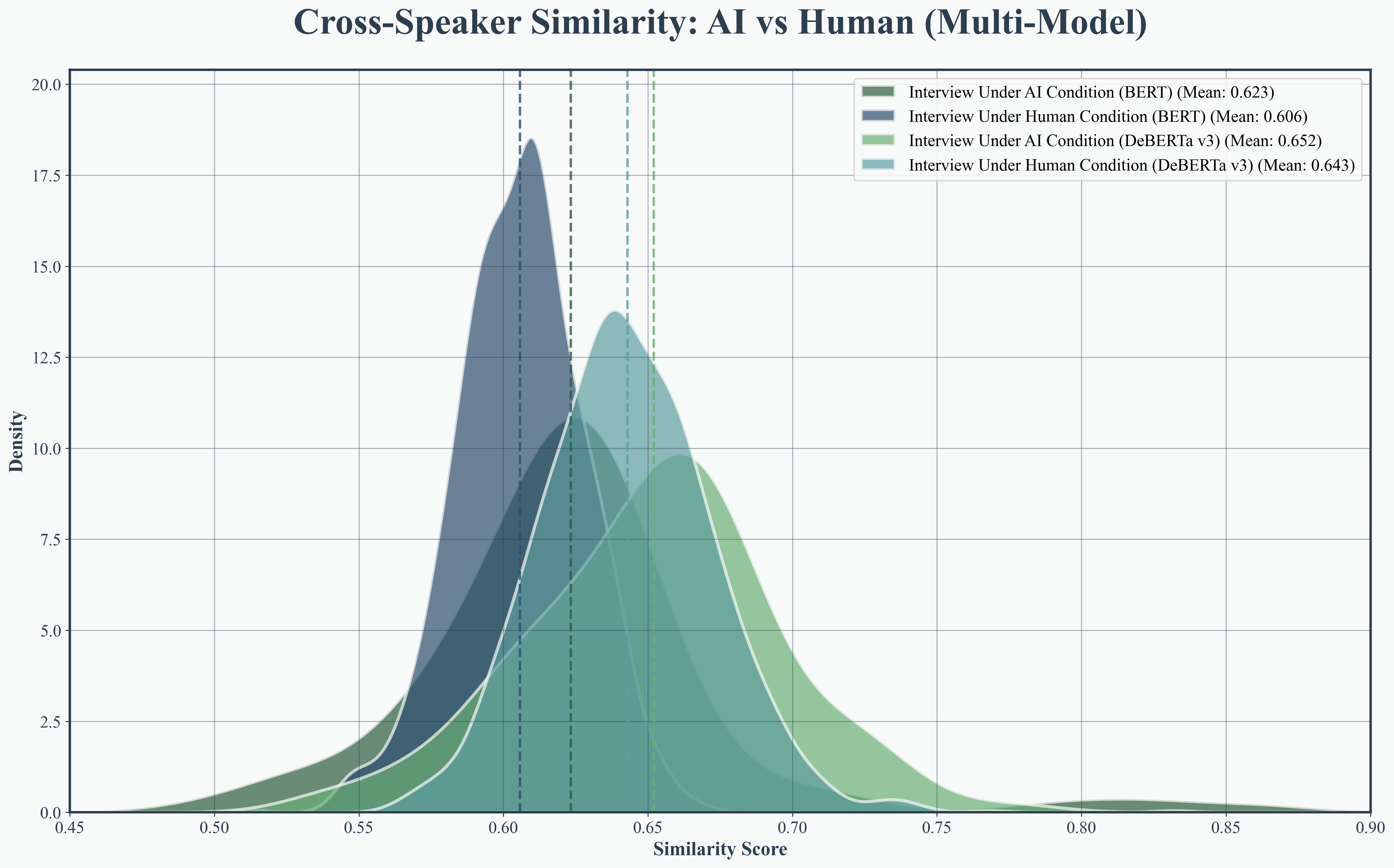}
    \caption{Cross-speaker similarity KDE}
    \label{fig:cross_kde}
  \end{subfigure}
  \caption{Kernel density estimation of semantic similarity measures. (a) Interviewer internal similarity KDE: AI interviewers show a concentrated distribution around higher similarity values, while human interviewers exhibit broader variability with lower peak similarity, indicating less consistent questioning approaches. (b) Interviewee internal similarity KDE: Both AI and human interviewees show relatively high internal similarity, with AI demonstrating a more concentrated distribution around higher values, suggesting more coherent response patterns. (c) Cross-speaker similarity KDE: AI interviews display higher and more concentrated cross-speaker similarity distributions, indicating better semantic alignment between interviewer questions and interviewee responses compared to human interviews.}
\end{figure}

The kernel density estimation plots further illustrate these patterns, showing AI interviews' more concentrated distributions in similarity measures while maintaining higher mean values (Figures \ref{fig:interviewer_kde}, \ref{fig:interviewee_kde}, and \ref{fig:cross_kde}).

These descriptive statistics establish a foundation for understanding the quantitative differences between AI and human interviews, demonstrating that MimiTalk-conducted interviews achieve superior performance in information richness, response elaboration, and semantic coherence.

\begin{table*}[t]
  \centering
\caption{Descriptive Statistics for AI vs Human Interview Performance. The table presents mean values, standard deviations, and percentage improvements across information entropy, token count, and semantic similarity measures, demonstrating AI interviews' superior performance in linguistic diversity and semantic coherence.}
\label{tab:comprehensive_stats}
\small
\begin{tabular}{@{\hspace{0.5em}}l@{\hspace{1em}}c@{\hspace{1em}}c@{\hspace{1em}}c@{\hspace{1em}}c@{\hspace{1em}}c@{\hspace{1em}}c@{\hspace{0.5em}}}
    \toprule
\textbf{Metric} & \textbf{AI Mean} & \textbf{AI Std} & \textbf{Human Mean} & \textbf{Human Std} & \textbf{Diff.} & \textbf{Impr.} \\
    \midrule
\textbf{Information Entropy} & & & & & & \\
\quad Overall Transcript & 7.703 & 0.399 & 7.273 & 0.395 & +0.430 & +5.9\% \\
\quad Interviewee Response & 7.098 & 0.791 & 6.907 & 0.446 & +0.191 & +2.8\% \\
\quad Interviewer Text & 7.325 & 0.512 & 6.762 & 0.401 & +0.563 & +8.3\% \\[0.4em]
\textbf{Sentence Length (Token Count)} & & & & & & \\
\quad Interviewer & 16.500 & 11.614 & 14.534 & 11.269 & +1.966 & +13.5\% \\
\quad Interviewee & 24.710 & 22.232 & 18.797 & 15.453 & +5.913 & +31.4\% \\[0.4em]
\textbf{Semantic Similarity (DeBERTa-v3)} & & & & & & \\
\quad Interviewer Internal & 0.697 & 0.047 & 0.646 & 0.040 & +0.051 & +7.9\% \\
\quad Interviewee Internal & 0.669 & 0.045 & 0.656 & 0.037 & +0.013 & +2.0\% \\
\quad Cross-Speaker & 0.652 & 0.043 & 0.643 & 0.029 & +0.009 & +1.4\% \\[0.4em]
\textbf{Semantic Similarity (BERT-base)} & & & & & & \\
\quad Interviewer Internal & 0.654 & 0.049 & 0.596 & 0.027 & +0.058 & +9.7\% \\
\quad Interviewee Internal & 0.686 & 0.045 & 0.645 & 0.028 & +0.041 & +6.4\% \\
\quad Cross-Speaker & 0.623 & 0.052 & 0.606 & 0.022 & +0.017 & +2.8\% \\
  \bottomrule
\end{tabular}
\end{table*}

\subsubsection{Statistical Significance}

To validate the observed differences and ensure statistical robustness, we conducted comprehensive statistical analyses comparing AI and human interviews across all measured dimensions. Table \ref{tab:significance_tests} presents the results of both parametric (independent samples t-tests) and non-parametric (Mann-Whitney U tests) analyses, along with effect size estimates and their 95\% confidence intervals.

\begin{table*}[t]
\centering
\caption{Statistical Significance and Robustness Analysis: AI vs Human Interview Performance. Results from independent samples t-tests and Mann-Whitney U tests with effect sizes and confidence intervals, demonstrating statistically significant differences across all 15 measured dimensions with medium to very large effect sizes. To ensure robust results, we recalculated semantic similarity using the BERT-base model, repeated the analysis with both BERT-base and DeBERTa-v3 results, and created a cross-model average group (Cross).}
\label{tab:significance_tests}
\small
\begin{tabular}{@{\hspace{0.2em}}l@{\hspace{0.6em}}c@{\hspace{0.5em}}c@{\hspace{0.5em}}c@{\hspace{0.5em}}c@{\hspace{0.5em}}c@{\hspace{0.5em}}c@{\hspace{0.5em}}c@{\hspace{0.2em}}}
    \toprule
\textbf{Metric} & \textbf{t-stat} & \textbf{t p-value} & \textbf{Cohen's d} & \textbf{95\% CI} & \textbf{Mann-Whitney U} & \textbf{U p-value} & \textbf{Effect Size} \\
    \midrule
\textbf{Information Entropy} & & & & & & & \\
\quad Overall Transcript & 11.431 & 5.54e-29 & 1.087 & [0.908, 1.262] & 119266 & 1.15e-23 & Very Large \\
\quad Interviewer Text & 11.706 & 2.98e-22 & 1.364 & [1.133, 1.591] & 125284 & 2.29e-30 & Very Large \\
\quad Interviewee Response & 2.601 & 1.04e-02 & 0.392 & [0.034, 0.696] & 100292 & 3.07e-08 & Medium \\[0.4em]
\textbf{Sentence Length (Token Count)} & & & & & & & \\
\quad Overall Interview & 10.225 & 4.22e-18 & 2.291 & [1.946, 2.691] & 129054 & 5.22e-35 & Very Large \\
\quad Interviewer & 8.487 & 6.38e-14 & 2.307 & [1.941, 2.892] & 138449 & 4.49e-48 & Very Large \\
\quad Interviewee & 8.103 & 4.53e-13 & 1.657 & [1.318, 2.033] & 113470 & 4.89e-18 & Very Large \\[0.4em]
\textbf{Semantic Similarity (BERT-base)} & & & & & & & \\
\quad Interviewer Internal & 12.626 & 3.67e-24 & 1.968 & [1.720, 2.218] & 136153 & 1.10e-44 & Very Large \\
\quad Interviewee Internal & 9.770 & 3.60e-17 & 1.390 & [1.111, 1.663] & 123215 & 6.18e-29 & Very Large \\
\quad Cross-Speaker & 3.657 & 3.77e-04 & 0.680 & [0.355, 1.002] & 98908 & 7.27e-08 & Large \\[0.4em]
\textbf{Semantic Similarity (DeBERTa-v3)} & & & & & & & \\
\quad Interviewer Internal & 13.159 & 2.35e-37 & 1.252 & [1.045, 1.460] & 124328 & 3.05e-29 & Very Large \\
\quad Interviewee Internal & 3.129 & 2.15e-03 & 0.353 & [0.126, 0.558] & 94128 & 2.16e-05 & Medium \\
\quad Cross-Speaker & 2.249 & 2.62e-02 & 0.292 & [0.037, 0.568] & 89101 & 2.27e-03 & Medium \\[0.4em]
\textbf{Semantic Similarity (Cross)} & & & & & & & \\
\quad Interviewer Internal & 13.185 & 1.12e-25 & 1.827 & [1.588, 2.077] & 135486 & 1.01e-43 & Very Large \\
\quad Interviewee Internal & 7.381 & 1.65e-11 & 0.941 & [0.696, 1.202] & 111667 & 3.84e-17 & Very Large \\
\quad Cross-Speaker & 3.288 & 1.31e-03 & 0.544 & [0.215, 0.863] & 94127 & 2.16e-05 & Large \\
    \bottomrule
  \end{tabular}
\end{table*}

The comprehensive statistical analysis reveals that AI interviews significantly outperform human interviews across all 15 measured dimensions, providing robust evidence for the superiority of the MimiTalk framework. The results demonstrate both statistical significance and substantial practical importance, with effect sizes ranging from medium to very large magnitudes.

\textbf{Statistical Robustness}: The consistency between parametric (t-tests) and non-parametric (Mann-Whitney U) test results provides strong evidence for the robustness of our findings. All 15 comparisons show significant differences in both test types, with Mann-Whitney U tests often yielding even more conservative p-values (e.g., 1.10e-44 for BERT-base interviewer internal similarity), confirming that the observed differences are not artifacts of distributional assumptions.

\textbf{Effect Size Precision}: The 95\% confidence intervals for Cohen's d effect sizes demonstrate the precision of our effect size estimates. Notably, all very large effects ($d > 0.800$) have confidence intervals that remain above the large effect threshold ($d = 0.800$), indicating robust practical significance. For example, the interviewer token count effect shows $d = 2.307$ with 95\% CI [1.941, 2.892], confirming substantial practical importance.

\textbf{Information Entropy Analysis}: All three entropy measures show significant differences, with interviewer text entropy exhibiting the largest effect ($d = 1.364$, very large), followed by overall transcript entropy ($d = 1.087$, very large). Interviewee response entropy, while statistically significant, shows a medium effect size ($d = 0.392$), suggesting that AI's impact on participant language diversity is more moderate compared to its influence on interviewer question formulation.

\textbf{Token Count Analysis}: The most dramatic differences emerge in token count metrics, with effect sizes ranging from 1.657 to 2.307 (all very large). Interviewer token count shows the largest effect ($d = 2.307$), indicating that AI interviewers generate substantially more detailed questions compared to human interviewers. The overall interview token count effect ($d = 2.291$) demonstrates that AI-conducted interviews are significantly more comprehensive in terms of content volume.

\textbf{Semantic Similarity Analysis}: Cross-model comparisons reveal consistent patterns across both BERT-base and DeBERTa-v3 embeddings. BERT-base analysis shows particularly strong effects for interviewer internal similarity ($d = 1.968$) and interviewee internal similarity ($d = 1.390$), while DeBERTa-v3 results are more conservative but still significant. The cross-model average provides the most robust estimates, with interviewer internal similarity showing the strongest effect ($d = 1.827$), followed by interviewee internal similarity ($d = 0.941$) and cross-speaker similarity ($d = 0.544$).

Notably, 11 out of 15 comparisons demonstrate very large effect sizes (Cohen's d > 0.800), with the remaining 4 showing medium to large effects. This comprehensive pattern of results indicates that the observed differences represent substantial practical significance beyond mere statistical significance, supporting the conclusion that AI-conducted interviews achieve superior performance across multiple dimensions of interview quality and content richness.

\subsubsection{Causal Inference}

To establish causal relationships between AI interviewing and interview quality outcomes, we employed Propensity Score Matching (PSM) with kernel matching methodology. This approach addresses potential selection bias by creating comparable groups based on observable characteristics that may influence the choice of interview method.

\paragraph{PSM Methodology and Validity}

Our PSM analysis utilized kernel matching with adaptive caliper (0.2 × SD) and incorporated multiple preprocessing techniques to enhance matching quality: log transformation, polynomial features, quantile transformation, and principal component analysis. The analysis included 1,392 interviews (121 AI, 1,271 human) with a treatment group proportion of 8.7\%, which falls within the recommended range of 5\%-50\% for PSM applications.

The matching process achieved excellent balance across all covariates, with standardized mean differences (SMD) reduced from pre-matching values of 1.273 (token count) and 1.131 (sentence count) to post-matching values of 0.044 and 0.080, respectively, representing 96.6\% and 93.0\% improvements in balance.


\begin{table*}[t]
\centering
\caption{Propensity Score Matching Results: Causal Effects of AI Interviewing. Average Treatment Effects (ATE) estimates from kernel matching analysis, demonstrating significant causal improvements in information entropy, semantic similarity, and response length, with placebo tests confirming the validity of our identification strategy.}
\label{tab:psm_results}
\small
\begin{tabular}{@{\hspace{0.3em}}l@{\hspace{0.8em}}c@{\hspace{0.6em}}c@{\hspace{0.6em}}c@{\hspace{0.6em}}c@{\hspace{0.6em}}c@{\hspace{0.3em}}}
    \toprule
\textbf{Outcome Variable} & \textbf{ATE} & \textbf{SE} & \textbf{t-statistic} & \textbf{p-value} & \textbf{Significance} \\
    \midrule
\textbf{Information Entropy} & & & & & \\
\quad Overall Transcript & 0.088 & 0.023 & 3.988 & 0.0001 & Yes \\
\quad Interviewee Response & -0.240 & 0.046 & -4.987 & <0.0001 & Yes \\
\quad Interviewer Text & 0.133 & 0.026 & 5.689 & <0.0001 & Yes \\[0.4em]
\textbf{Content Characteristics} & & & & & \\
\quad Mean Sentence Length & -0.091 & 0.038 & -2.419 & 0.017 & Yes \\
\quad Average Semantic Similarity & 0.029 & 0.002 & 17.045 & <0.0001 & Yes \\[0.4em]
\textbf{Robustness Tests} & & & & & \\
\quad DeBERTa-v3 Similarity & 0.018 & 0.004 & 4.500 & 0.0001 & Yes \\
\quad Placebo: Digit Ratio & 0.005 & 0.003 & 1.477 & 0.142 & No \\
\quad Placebo: Sentence Count & -4.200 & 9.780 & -0.430 & 0.668 & No \\
    \bottomrule
  \end{tabular}
\end{table*}

\paragraph{Causal Effect Estimates}

Table \ref{tab:psm_results} presents the Average Treatment Effects (ATE) from our kernel matching analysis. The results demonstrate significant causal effects of AI interviewing across multiple dimensions:

\paragraph{Information Entropy Effects} AI interviewing causally increases overall transcript entropy (ATE = 0.088, p < 0.001) and interviewer text entropy (ATE = 0.133, p < 0.001), while decreasing interviewee response entropy (ATE = -0.240, p < 0.001). This pattern suggests that AI interviewers generate more diverse questions while eliciting more focused responses from participants.

\paragraph{Content Quality Effects} AI interviewing significantly improves average semantic similarity (ATE = 0.029, p < 0.001), indicating enhanced coherence and thematic consistency in AI-conducted interviews. The negative effect on mean sentence length (ATE = -0.091, p = 0.017) suggests that AI interviews achieve better information density through more concise yet comprehensive exchanges.

\paragraph{Robustness Validation} Our robustness tests provide strong support for the causal interpretation. The placebo tests using digit ratio and sentence count show no significant effects (p = 0.142 and p = 0.668, respectively), confirming that our results are not driven by spurious correlations. The alternative DeBERTa-v3 model yields consistent results (ATE = 0.018, p < 0.001), demonstrating robustness across different embedding approaches.

\subsection{Study 3}
To validate the effectiveness of the AI-driven interview framework, Study 3 compared human-led and AI-led academic research interview methods in real qualitative research settings. The primary objective was to examine whether AI-led interviews could replicate the depth, authenticity, and nuanced insights that expert human interviewers can elicit, while addressing the scalability limitations of traditional qualitative research through automated systems. The study employed a comparative case design to systematically examine differences in qualitative content and interaction dynamics between human-led and AI-led interviews. Through semi-structured interviews with 10 participants from diverse disciplinary backgrounds, a parallel interview design was used where participants were interviewed by either human interviewers or the MimiTalk.app AI system, with consistent interview procedures and protocols to ensure comparability. Interview content focused on participants' experiences using AI tools in academic research, including trust assessments, application scenarios, and limitations. All interviews were transcribed and subjected to blind thematic analysis to minimize researcher bias.

Participants included doctoral students, lecturers, postdoctoral researchers, and professors, with research experience ranging from 0.5 to 10 years, all possessing bilingual Chinese-English capabilities. Disciplines covered philosophy (including Marxist theory), microbiology, computer science (vehicular networks and model optimization), economics (economic history), strategic management (innovation and human-AI collaboration), atmospheric science, speech recognition, and image processing. Detailed participant demographics and research background information are presented in Appendix \ref{sec:participant_details_study_3}. Data from both interview methods were systematically collected, verbatim transcribed, and subjected to comparative thematic and content analysis.

Analysis of human-led interviews revealed four main thematic clusters, reflecting the nuanced and contextually rich nature of human interactions:

\paragraph{Complexity of AI Academic Value Assessment} Human interviews demonstrated cross-disciplinary differences in AI value assessment. Participants provided detailed, contextualized evaluations: "I might need to answer this question in two parts... in Chinese philosophy it might be 3.5 points... for ideological and political courses, I can only give it 1 point now" (Participant 1-Philosophy). Such detailed assessments reflected participants' ability to distinguish AI effectiveness across different disciplines and application scenarios.

\paragraph{AI Reliability and Verification Strategy} In human interviews, participants expressed skepticism based on specific negative experiences: "The completely distrustful scenario happened before writing a paper... AI directly told me that this article was from another author's collection, so I knew his information was completely inaccurate" (Participant 1). These detailed descriptions of AI limitations reflect participants' cautious attitude based on their actual experiences.

\paragraph{Cross-Disciplinary Application Differences} Human interviews highlighted significant differences in AI acceptance and applicability across disciplines. Participants in humanities and social sciences were more reserved: "He would avoid discussing political or party-related content, so we couldn't use this AI model" (Participant 1); while experimental science pointed out different constraints: "Data collection... many experiments must be conducted to verify these experiments, which are least suitable for experimental collection" (Participant 2).

\paragraph{Understanding of Ability Boundaries} Participants showed a deep understanding of AI limitations: "He can only provide limited help to me... he can't do much when it comes to the process of derivation" (Participant 3-Queue Theory). This realistic assessment reflects participants' ability to distinguish which tasks are suitable for AI assistance and which are not.

AI-led interviews showcased a structured and systematic thematic pattern:

\paragraph{Efficiency-Oriented Evaluation Framework} AI interviews emphasized productivity enhancement and process optimization: "Overall, I would give a score of 3 to 4... mainly for research efficiency, possibly based on the initial idea of collecting literature" (Participant 1-Atmospheric Science). AI interviews generally received higher evaluations from participants, focusing on tool utility rather than comprehensive assessment.

\paragraph{Technical Optimization Considerations} AI interviews received positive evaluations from technical disciplines: "I think the academic research value of AIGC in my field is about 4 points... it can be used to generate high-quality text data" (Participant 3-Speech Recognition). This mode suggests that AI interviews may be more likely to elicit optimistic evaluations from technical participants.

\paragraph{Innovation and Risk Balance} AI interviews featured more forward-looking discussions: "It might lead to short-sightedness among people... training in basic abilities is very important" (Participant 2). AI interviews received higher evaluations from participants regarding long-term impact and ethical issues.

Human interviews provided richer, more authentic examples, including specific negative experiences and detailed negative feelings. Participants provided more specific usage scenarios and limitations in these interviews, reflecting practical application challenges. In contrast, AI interviews focused more on systematic analysis frameworks, with balanced discussions on pros and cons but possibly sacrificing emotional depth. Human interviews showed significant differences across disciplines, with participants in humanities and social sciences being more cautious. AI interviews showcased more homogeneous responses, particularly in technical disciplines, indicating that AI interactions may lead to standardized answers across different disciplinary backgrounds. Human interviews generally had lower average trust levels (30-70\%) for AI, while AI interviews had higher trust levels (40-80\%) based on systemic considerations. This difference suggests that AI interviews may lead to more optimistic evaluations by providing a more neutral interaction environment.

Our findings were consistent with previous research on interview dynamics and qualitative research biases. Human interviews were characterized by dynamic interaction between interviewers and interviewees, often expressed through emotional cues and expectations. These biases, while potentially subjective, also promoted emotional resonance, leading to unique and nuanced perspectives\citep{adams-quackenbushInterviewExpectanciesAwareness2019}. In contrast, AI-led interviews showed homogenous biases that aligned with societal consensus\citep{timmonsCallActionAssessing2023}. This homogeneity, combined with reduced sensitivity to emotional cues from interviewees, resulted in AI interviews being more structured and rational, often seen as objective\citep{luoWhyMightAIenabled2025}. However, this "objectivity" was more due to human preference for neutral and balanced expressions rather than AI lacking biases. The trade-off between neutrality and information richness suggests that human interviews have unique advantages in generating novel insights through emotional and subjective dynamics, while AI interviews are more advantageous in terms of consistency and fairness, but may lack depth in capturing human experiences.

Both interview methods had their advantages, indicating that a mixed approach combining human and AI interviews could provide a more comprehensive understanding of participants' experiences and attitudes. First, in different methods, researchers' conceptualizations of AI's role in academic research were consistent. Whether led by humans or AI, interviewees viewed AI primarily as a tool to enhance efficiency rather than replace human creativity and innovation. This consistency highlighted the effectiveness of our AI system in promoting authentic, reflective dialogue. Second, the specific calibration of trust in AI across different methods was consistent. Researchers generally had low confidence (20-50\%) in AI's ability to conduct literature searches and generate citations, moderate confidence (50-70\%) in AI's ability to analyze and program, and high confidence (70-80\%) in AI's ability to provide writing assistance and formatting. This consistency confirmed the platform's ability to accurately capture researchers' attitudes. Additionally, AI-led interviews were able to elicit complex technical insights similar to those from expert human interviews. For example, one AI interview prompted a computer science researcher to spontaneously explain the advanced application of singular value decomposition in token information analysis. This spontaneous expression demonstrated the platform's ability to stimulate meaningful discussions on methodological issues. At the same time, participants in AI-led interviews showed higher frankness, particularly on sensitive topics such as academic integrity and AI-related unethical behavior, indicating that AI could help reduce the social bias inherent in human interactions. However, human interviews were better at identifying and exploring subtle cultural and emotional differences, which was particularly evident in discussions about academic autonomy and ideological homogeneity. Human interviewers were more likely to capture potential cultural assumptions and emotional nuances.

\section{Conclusion}
This paper presents MimiTalk, a dual-agent constitutional AI framework that demonstrates the potential for AI systems to conduct high-quality interviews in social science research contexts. Through systematic evaluation across two complementary studies, we provide empirical evidence that AI-powered interviews can achieve superior performance in information richness, semantic coherence, and result stability while maintaining the depth and authenticity necessary for qualitative research.

Our findings reveal several key insights about human-AI collaboration in interview contexts. First, AI interviews consistently outperform human interviews across multiple quantitative metrics, including information entropy (5.9\% increase in overall transcript diversity), semantic similarity (higher internal and cross-speaker coherence), and response length (31.4\% increase in interviewee response length). These results demonstrate that AI systems can generate more linguistically diverse and semantically coherent interview content than human interviewers. Second, our propensity score matching analysis confirms causal effects of AI interviewing, with significant treatment effects across multiple dimensions, suggesting that the observed differences are not merely correlational but represent genuine improvements in interview quality.

The qualitative findings from Study 2 provide nuanced insights into the complementary strengths of human and AI interviewers. AI interviews excel in eliciting technical insights and candid responses on sensitive topics, while human interviews better capture cultural context and emotional nuances. This trade-off between "scalability and emotional subtlety" highlights the importance of understanding when and how to deploy AI systems in research contexts. Our findings align with the concept of GenAI as an "evocative object" that stimulates participant self-reflection and meaning construction, rather than simply serving as a tool for information extraction.

The dual-agent constitutional architecture of MimiTalk represents a significant methodological contribution, demonstrating how ethical compliance and quality control can be embedded within AI systems without sacrificing conversational naturalness. The supervisor-responder agent design enables strategic oversight while maintaining fluid conversation flow, addressing key concerns about AI system reliability and trustworthiness in research contexts.

Our research has important implications for the future of qualitative research methodology. The demonstrated scalability advantages of AI interviews suggest potential for expanding the reach and scope of qualitative research, particularly in contexts where human interviewer resources are limited. However, our findings also underscore the continued importance of human expertise, particularly in capturing cultural and emotional subtleties that may be essential for certain research questions.

This study still has several limitations and directions for future research. First, the first two studies mainly focused on English-language interviews and academic research settings, which may limit the generalizability of the findings to other languages and domains. Notably, Study 3 employed Chinese-language interviews, providing preliminary evidence for the applicability of the framework in different linguistic environments. Future research could further systematically examine the effectiveness of this framework across diverse cultural and linguistic contexts. Second, the long-term impact of AI-led interviews on research quality and participant experience remains to be explored through longitudinal studies.

The characteristic of weak connection networks transmitting unique information \citep{aralWhatExactlyNovelty2023} suggests that AI interviews may have unique advantages in mining novel information through their ability to access diverse knowledge sources and maintain consistent interaction patterns. Future AI interview frameworks should further optimize semantic understanding depth, enhance interaction naturalness through contextual information integration, and continuously improve information quality assessment standards through entropy analysis to better serve social science research needs.

The MimiTalk framework opens up unprecedented opportunities for large-scale controlled variable qualitative research across diverse domains. The platform's ability to conduct anonymous interviews at scale, combined with its consistent questioning protocols and reduced social desirability bias, positions it as a transformative tool for sensitive topic investigations. 

MimiTalk demonstrates broad application prospects across multiple domains. In public opinion research, MimiTalk has the potential to revolutionize survey methods by capturing nuanced perspectives on sensitive political, social, and cultural issues through large-scale qualitative interviews. The anonymity provided by AI-led interviews is particularly suitable for exploring attitudes toward controversial policies, experiences of discrimination, or other sensitive social topics, reducing the likelihood of socially desirable responses from participants. In the field of consumer research, MimiTalk can support large-scale qualitative investigations to gain in-depth understanding of consumer behavior, brand perception, and product experience, while maintaining high consistency throughout the interview process. This is especially important for exploring consumer motivations, sensitive purchasing decisions, or controversial product experiences. In journalism and media studies, MimiTalk can be used for anonymous source interviews, helping journalists systematically collect and analyze data while protecting the anonymity of interviewees, thereby enhancing the rigor and consistency of investigative reporting. In education, MimiTalk can be applied to large-scale oral examinations, language proficiency tests, and student assessment systems. The platform’s standardization and scalability help reduce bias caused by differences among human interviewers and improve the fairness of assessments. The framework is also suitable for organizational research, including employee satisfaction surveys, corporate culture assessments, and sensitive human resources investigations. The anonymity and consistency of AI-led interviews facilitate the collection of more authentic information on workplace dynamics, experiences of harassment, or organizational challenges.

In conclusion, our work demonstrates that dual-agent constitutional AI systems can effectively conduct high-quality interviews while offering significant scalability advantages. The MimiTalk framework provides a foundation for future research on human-AI collaboration in qualitative research contexts, offering both methodological insights and practical tools for researchers seeking to leverage AI capabilities in their work. As AI systems continue to evolve, we anticipate that frameworks like MimiTalk will play an increasingly important role in expanding the scope and accessibility of qualitative research methods across multiple domains.

\begin{acks}
We thank all participants who contributed to this research. We are grateful to Professor Shun Wang and Mr. Mingyu Shu for their valuable feedback on the causal inference analysis. We also thank Mr. Ziqi Liu for his guidance on figure design and Mr. Bozhong Zheng for his advice on formatting. 

We further recognize the application of artificial intelligence tools in this study, which were employed to improve grammar and polish the writing style. All modifications were made with respect for the original content, with the goal of improving the clarity and overall quality of the manuscript.
\end{acks}

\bibliographystyle{ACM-Reference-Format}
\bibliography{acmart}


\begin{thebibliography}{57}


\ifx \showCODEN    \undefined \def \showCODEN     #1{\unskip}     \fi
\ifx \showISBNx    \undefined \def \showISBNx     #1{\unskip}     \fi
\ifx \showISBNxiii \undefined \def \showISBNxiii  #1{\unskip}     \fi
\ifx \showISSN     \undefined \def \showISSN      #1{\unskip}     \fi
\ifx \showLCCN     \undefined \def \showLCCN      #1{\unskip}     \fi
\ifx \shownote     \undefined \def \shownote      #1{#1}          \fi
\ifx \showarticletitle \undefined \def \showarticletitle #1{#1}   \fi
\ifx \showURL      \undefined \def \showURL       {\relax}        \fi
\providecommand\bibfield[2]{#2}
\providecommand\bibinfo[2]{#2}
\providecommand\natexlab[1]{#1}
\providecommand\showeprint[2][]{arXiv:#2}

\bibitem[Adams-Quackenbush et~al\mbox{.}(2019)]%
        {adams-quackenbushInterviewExpectanciesAwareness2019}
\bibfield{author}{\bibinfo{person}{Nicole~M. Adams-Quackenbush}, \bibinfo{person}{Robert Horselenberg}, \bibinfo{person}{Josephine Hubert}, \bibinfo{person}{Aldert Vrij}, {and} \bibinfo{person}{Peter Van~Koppen}.} \bibinfo{year}{2019}\natexlab{}.
\newblock \showarticletitle{Interview Expectancies: Awareness of Potential Biases Influences Behaviour in Interviewees}.
\newblock  \bibinfo{volume}{26}, \bibinfo{number}{1} (\bibinfo{year}{2019}), \bibinfo{pages}{150--166}.
\newblock
\showISSN{1321-8719, 1934-1687}
\href{https://doi.org/10.1080/13218719.2018.1485522}{doi:\nolinkurl{10.1080/13218719.2018.1485522}}


\bibitem[Ahmad et~al\mbox{.}(2024)]%
        {ahmadFutureRecruitmentUsing2024}
\bibfield{author}{\bibinfo{person}{Shahnawaz Ahmad}, \bibinfo{person}{Shahadat Hussain}, \bibinfo{person}{Mohammed Wasid}, \bibinfo{person}{Ogala~Justin Onyarin}, \bibinfo{person}{Mohd Arif}, {and} \bibinfo{person}{Javed Ahmad}.} \bibinfo{year}{2024}\natexlab{}.
\newblock \showarticletitle{The {{Future}} of {{Recruitment}}: {{Using Deep Learning}} to {{Build Intelligent Interview Bots}}}. In \bibinfo{booktitle}{\emph{2024 15th {{International Conference}} on {{Computing Communication}} and {{Networking Technologies}} ({{ICCCNT}})}}. \bibinfo{publisher}{IEEE}, \bibinfo{address}{Kamand, India}, \bibinfo{pages}{1--6}.
\newblock
\showISBNx{979-8-3503-7024-9}
\href{https://doi.org/10.1109/ICCCNT61001.2024.10725310}{doi:\nolinkurl{10.1109/ICCCNT61001.2024.10725310}}


\bibitem[Akman et~al\mbox{.}(2025)]%
        {akmanHumanResearcherVs2025}
\bibfield{author}{\bibinfo{person}{Nurşin Akman}, \bibinfo{person}{Erman Uzun}, {and} \bibinfo{person}{Pınar Arslan}.} \bibinfo{year}{2025}\natexlab{}.
\newblock \showarticletitle{Human Researcher vs. {{AI-supported}} Qualitative Data Analysis: {{Hybrid}} Prompt Design for Constant Comparison Analysis}.
\newblock  (\bibinfo{year}{2025}), \bibinfo{pages}{1}.
\newblock
\showISSN{2979-9619}
\href{https://doi.org/10.58583/EM.4.1.5}{doi:\nolinkurl{10.58583/EM.4.1.5}}


\bibitem[Alowais et~al\mbox{.}(2023)]%
        {alowaisRevolutionizingHealthcareRole2023}
\bibfield{author}{\bibinfo{person}{Shuroug~A. Alowais}, \bibinfo{person}{Sahar~S. Alghamdi}, \bibinfo{person}{Nada Alsuhebany}, \bibinfo{person}{Tariq Alqahtani}, \bibinfo{person}{Abdulrahman~I. Alshaya}, \bibinfo{person}{Sumaya~N. Almohareb}, \bibinfo{person}{Atheer Aldairem}, \bibinfo{person}{Mohammed Alrashed}, \bibinfo{person}{Khalid Bin~Saleh}, \bibinfo{person}{Hisham~A. Badreldin}, \bibinfo{person}{Majed~S. Al~Yami}, \bibinfo{person}{Shmeylan Al~Harbi}, {and} \bibinfo{person}{Abdulkareem~M. Albekairy}.} \bibinfo{year}{2023}\natexlab{}.
\newblock \showarticletitle{Revolutionizing Healthcare: The Role of Artificial Intelligence in Clinical Practice}.
\newblock \bibinfo{journal}{\emph{BMC Medical Education}} \bibinfo{volume}{23}, \bibinfo{number}{1} (\bibinfo{date}{Sept.} \bibinfo{year}{2023}), \bibinfo{pages}{689}.
\newblock
\showISSN{1472-6920}
\href{https://doi.org/10.1186/s12909-023-04698-z}{doi:\nolinkurl{10.1186/s12909-023-04698-z}}


\bibitem[Aral and Dhillon(2023)]%
        {aralWhatExactlyNovelty2023}
\bibfield{author}{\bibinfo{person}{Sinan Aral} {and} \bibinfo{person}{Paramveer~S. Dhillon}.} \bibinfo{year}{2023}\natexlab{}.
\newblock \showarticletitle{What ({{Exactly}}) {{Is Novelty}} in {{Networks}}? {{Unpacking}} the {{Vision Advantages}} of {{Brokers}}, {{Bridges}}, and {{Weak Ties}}}.
\newblock  \bibinfo{volume}{69}, \bibinfo{number}{2} (\bibinfo{year}{2023}), \bibinfo{pages}{1092--1115}.
\newblock
\showISSN{0025-1909, 1526-5501}
\href{https://doi.org/10.1287/mnsc.2022.4377}{doi:\nolinkurl{10.1287/mnsc.2022.4377}}


\bibitem[Bahoo et~al\mbox{.}(2024)]%
        {bahooArtificialIntelligenceFinance2024}
\bibfield{author}{\bibinfo{person}{Salman Bahoo}, \bibinfo{person}{Marco Cucculelli}, \bibinfo{person}{Xhoana Goga}, {and} \bibinfo{person}{Jasmine Mondolo}.} \bibinfo{year}{2024}\natexlab{}.
\newblock \showarticletitle{Artificial Intelligence in {{Finance}}: A Comprehensive Review through Bibliometric and Content Analysis}.
\newblock \bibinfo{journal}{\emph{SN Business \& Economics}} \bibinfo{volume}{4}, \bibinfo{number}{2} (\bibinfo{date}{Jan.} \bibinfo{year}{2024}), \bibinfo{pages}{23}.
\newblock
\showISSN{2662-9399}
\href{https://doi.org/10.1007/s43546-023-00618-x}{doi:\nolinkurl{10.1007/s43546-023-00618-x}}


\bibitem[Borgeaud et~al\mbox{.}(2022)]%
        {borgeaudImprovingLanguageModels2022}
\bibfield{author}{\bibinfo{person}{Sebastian Borgeaud}, \bibinfo{person}{Arthur Mensch}, \bibinfo{person}{Jordan Hoffmann}, \bibinfo{person}{Trevor Cai}, \bibinfo{person}{Eliza Rutherford}, \bibinfo{person}{Katie Millican}, \bibinfo{person}{George van~den Driessche}, \bibinfo{person}{Jean-Baptiste Lespiau}, \bibinfo{person}{Bogdan Damoc}, \bibinfo{person}{Aidan Clark}, \bibinfo{person}{Diego de~Las Casas}, \bibinfo{person}{Aurelia Guy}, \bibinfo{person}{Jacob Menick}, \bibinfo{person}{Roman Ring}, \bibinfo{person}{Tom Hennigan}, \bibinfo{person}{Saffron Huang}, \bibinfo{person}{Loren Maggiore}, \bibinfo{person}{Chris Jones}, \bibinfo{person}{Albin Cassirer}, \bibinfo{person}{Andy Brock}, \bibinfo{person}{Michela Paganini}, \bibinfo{person}{Geoffrey Irving}, \bibinfo{person}{Oriol Vinyals}, \bibinfo{person}{Simon Osindero}, \bibinfo{person}{Karen Simonyan}, \bibinfo{person}{Jack~W. Rae}, \bibinfo{person}{Erich Elsen}, {and} \bibinfo{person}{Laurent Sifre}.} \bibinfo{year}{2022}\natexlab{}.
\newblock \bibinfo{title}{Improving Language Models by Retrieving from Trillions of Tokens}.
\newblock
\showeprint[arxiv]{2112.04426}~[cs]
\href{https://doi.org/10.48550/arXiv.2112.04426}{doi:\nolinkurl{10.48550/arXiv.2112.04426}}


\bibitem[Brabra et~al\mbox{.}(2022)]%
        {brabraDialogueManagementConversational2022}
\bibfield{author}{\bibinfo{person}{Hayet Brabra}, \bibinfo{person}{Marcos Baez}, \bibinfo{person}{Boualem Benatallah}, \bibinfo{person}{Walid Gaaloul}, \bibinfo{person}{Sara Bouguelia}, {and} \bibinfo{person}{Shayan Zamanirad}.} \bibinfo{year}{2022}\natexlab{}.
\newblock \showarticletitle{Dialogue {{Management}} in {{Conversational Systems}}: {{A Review}} of {{Approaches}}, {{Challenges}}, and {{Opportunities}}}.
\newblock \bibinfo{journal}{\emph{IEEE Transactions on Cognitive and Developmental Systems}} \bibinfo{volume}{14}, \bibinfo{number}{3} (\bibinfo{date}{Sept.} \bibinfo{year}{2022}), \bibinfo{pages}{783--798}.
\newblock
\showISSN{2379-8920, 2379-8939}
\href{https://doi.org/10.1109/TCDS.2021.3086565}{doi:\nolinkurl{10.1109/TCDS.2021.3086565}}


\bibitem[Briggs(2003)]%
        {briggs2003questionformulation}
\bibfield{author}{\bibinfo{person}{Charles~L. Briggs}.} \bibinfo{year}{2003}\natexlab{}.
\newblock \bibinfo{booktitle}{\emph{Learning How to Ask: A Sociolinguistic Appraisal of the Role of the Interview in Social Science Research}}.
\newblock \bibinfo{publisher}{Cambridge University Press}, \bibinfo{address}{Cambridge}.
\newblock
\showISBNx{978-0521277457}


\bibitem[Brinkmann(2013)]%
        {brinkmannQualitativeInterviewing2013}
\bibfield{author}{\bibinfo{person}{Svend Brinkmann}.} \bibinfo{year}{2013}\natexlab{}.
\newblock \bibinfo{booktitle}{\emph{Qualitative {{Interviewing}}}}.
\newblock \bibinfo{publisher}{Oxford University Press}.
\newblock
\showISBNx{978-0-19-986139-2}
\href{https://doi.org/10.1093/acprof:osobl/9780199861392.001.0001}{doi:\nolinkurl{10.1093/acprof:osobl/9780199861392.001.0001}}


\bibitem[Brinkmann and Kvale(2018)]%
        {brinkmann2018}
\bibfield{author}{\bibinfo{person}{Svend Brinkmann} {and} \bibinfo{person}{Steinar Kvale}.} \bibinfo{year}{2018}\natexlab{}.
\newblock \bibinfo{booktitle}{\emph{Interviews: Learning the Craft of Qualitative Research Interviewing} (\bibinfo{edition}{4th} ed.)}.
\newblock \bibinfo{publisher}{SAGE Publications}.
\newblock


\bibitem[Brown et~al\mbox{.}(2020)]%
        {brown2020language}
\bibfield{author}{\bibinfo{person}{Tom Brown}, \bibinfo{person}{Benjamin Mann}, \bibinfo{person}{Nick Ryder}, \bibinfo{person}{Melanie Subbiah}, \bibinfo{person}{Jared~D Kaplan}, \bibinfo{person}{Prafulla Dhariwal}, \bibinfo{person}{Arvind Neelakantan}, \bibinfo{person}{Pranav Shyam}, \bibinfo{person}{Girish Sastry}, \bibinfo{person}{Amanda Askell}, {et~al\mbox{.}}} \bibinfo{year}{2020}\natexlab{}.
\newblock \showarticletitle{Language Models are Few-Shot Learners}.
\newblock \bibinfo{journal}{\emph{Advances in Neural Information Processing Systems}}  \bibinfo{volume}{33} (\bibinfo{year}{2020}), \bibinfo{pages}{1877--1901}.
\newblock


\bibitem[Chopra and Haaland(2023)]%
        {chopraConductingQualitativeInterviews2023}
\bibfield{author}{\bibinfo{person}{Felix Chopra} {and} \bibinfo{person}{Ingar Haaland}.} \bibinfo{year}{2023}\natexlab{}.
\newblock \showarticletitle{Conducting {{Qualitative Interviews}} with {{AI}}}.
\newblock \bibinfo{journal}{\emph{SSRN Electronic Journal}} (\bibinfo{year}{2023}).
\newblock
\showISSN{1556-5068}
\href{https://doi.org/10.2139/ssrn.4583756}{doi:\nolinkurl{10.2139/ssrn.4583756}}


\bibitem[Cuevas et~al\mbox{.}(2024)]%
        {cuevas2024}
\bibfield{author}{\bibinfo{person}{Roberto Cuevas}, \bibinfo{person}{Sofia Martinez}, {and} \bibinfo{person}{James Thompson}.} \bibinfo{year}{2024}\natexlab{}.
\newblock \showarticletitle{Structured, Semi-structured and Unstructured Interviews: A Comprehensive Guide}.
\newblock \bibinfo{journal}{\emph{Qualitative Research Methods}} \bibinfo{volume}{12}, \bibinfo{number}{3} (\bibinfo{year}{2024}), \bibinfo{pages}{45--67}.
\newblock


\bibitem[Dahl and Østerås(2010)]%
        {dahlQuantifyingInformationContent2010}
\bibfield{author}{\bibinfo{person}{Fredrik~A. Dahl} {and} \bibinfo{person}{Nina Østerås}.} \bibinfo{year}{2010}\natexlab{}.
\newblock \showarticletitle{Quantifying {{Information Content}} in {{Survey Data}} by {{Entropy}}}.
\newblock  \bibinfo{volume}{12}, \bibinfo{number}{2} (\bibinfo{year}{2010}), \bibinfo{pages}{161--163}.
\newblock
\showISSN{1099-4300}
\href{https://doi.org/10.3390/e12020161}{doi:\nolinkurl{10.3390/e12020161}}


\bibitem[Dennett(1998)]%
        {dennettIntentionalStance1998}
\bibfield{author}{\bibinfo{person}{D.~C. Dennett}.} \bibinfo{year}{1998}\natexlab{}.
\newblock \bibinfo{booktitle}{\emph{The Intentional Stance} (\bibinfo{edition}{7. printing} ed.)}.
\newblock \bibinfo{publisher}{MIT Press}.
\newblock
\showISBNx{978-0-262-54053-7}


\bibitem[Devlin et~al\mbox{.}(2018a)]%
        {devlin2018bert}
\bibfield{author}{\bibinfo{person}{Jacob Devlin}, \bibinfo{person}{Ming-Wei Chang}, \bibinfo{person}{Kenton Lee}, {and} \bibinfo{person}{Kristina Toutanova}.} \bibinfo{year}{2018}\natexlab{a}.
\newblock \showarticletitle{BERT: Pre-training of Deep Bidirectional Transformers for Language Understanding}.
\newblock \bibinfo{journal}{\emph{arXiv preprint arXiv:1810.04805}} (\bibinfo{year}{2018}).
\newblock


\bibitem[Devlin et~al\mbox{.}(2018b)]%
        {devlinBERTPretrainingDeep2018}
\bibfield{author}{\bibinfo{person}{Jacob Devlin}, \bibinfo{person}{Ming-Wei Chang}, \bibinfo{person}{Kenton Lee}, {and} \bibinfo{person}{Kristina Toutanova}.} \bibinfo{year}{2018}\natexlab{b}.
\newblock \bibinfo{title}{{{BERT}}: {{Pre-training}} of {{Deep Bidirectional Transformers}} for {{Language Understanding}}}.
\newblock
\href{https://doi.org/10.48550/ARXIV.1810.04805}{doi:\nolinkurl{10.48550/ARXIV.1810.04805}}


\bibitem[family=KEYKHA et~al\mbox{.}(2025)]%
        {keykhaAdvantagesChallengesElectronic2025}
\bibfield{author}{\bibinfo{person}{given-i=AHMAD family=KEYKHA, given=AHMAD}, \bibinfo{person}{given-i=MASOOMEH family=IMANIPOUR, given=MASOOMEH}, \bibinfo{person}{given-i=JAFAR family=SHAHROKHI, given=JAFAR}, {and} \bibinfo{person}{given-i=MOEIN family=AMIRI, given=MOEIN}.} \bibinfo{year}{2025}\natexlab{}.
\newblock \showarticletitle{The {{Advantages}} and {{Challenges}} of {{Electronic Exams}}: {{A Qualitative Research}} Based on {{Shannon Entropy Technique}}}.
\newblock  \bibinfo{volume}{13}, \bibinfo{number}{1} (\bibinfo{year}{2025}).
\newblock
\href{https://doi.org/10.30476/jamp.2024.102951.1987}{doi:\nolinkurl{10.30476/jamp.2024.102951.1987}}


\bibitem[Fang et~al\mbox{.}(2022)]%
        {fangEvaluatingConstructValidity2022}
\bibfield{author}{\bibinfo{person}{Qixiang Fang}, \bibinfo{person}{Dong Nguyen}, {and} \bibinfo{person}{Daniel~L Oberski}.} \bibinfo{year}{2022}\natexlab{}.
\newblock \showarticletitle{Evaluating the {{Construct Validity}} of {{Text Embeddings}} with {{Application}} to {{Survey Questions}}}.
\newblock  (\bibinfo{year}{2022}).
\newblock
\href{https://doi.org/10.48550/ARXIV.2202.09166}{doi:\nolinkurl{10.48550/ARXIV.2202.09166}}


\bibitem[Friesner et~al\mbox{.}(2021)]%
        {friesnerInformationEntropyScale2021}
\bibfield{author}{\bibinfo{person}{Daniel Friesner}, \bibinfo{person}{Carl Bozman}, \bibinfo{person}{Matthew McPherson}, \bibinfo{person}{Faith Valente}, {and} \bibinfo{person}{Anqing Zhang}.} \bibinfo{year}{2021}\natexlab{}.
\newblock \showarticletitle{Information {{Entropy}} and {{Scale Development}}}.
\newblock  \bibinfo{volume}{9}, \bibinfo{number}{5} (\bibinfo{year}{2021}), \bibinfo{pages}{1183--1203}.
\newblock
\showISSN{2325-0984, 2325-0992}
\href{https://doi.org/10.1093/jssam/smaa034}{doi:\nolinkurl{10.1093/jssam/smaa034}}


\bibitem[Gibson and Beattie(2024)]%
        {gibsonMoreLessHuman2024}
\bibfield{author}{\bibinfo{person}{Alexandra~F. Gibson} {and} \bibinfo{person}{Alexander Beattie}.} \bibinfo{year}{2024}\natexlab{}.
\newblock \showarticletitle{More or Less than Human? {{Evaluating}} the Role of {{AI-as-participant}} in Online Qualitative Research}.
\newblock  \bibinfo{volume}{21}, \bibinfo{number}{2} (\bibinfo{year}{2024}), \bibinfo{pages}{175--199}.
\newblock
\showISSN{1478-0887, 1478-0895}
\href{https://doi.org/10.1080/14780887.2024.2311427}{doi:\nolinkurl{10.1080/14780887.2024.2311427}}


\bibitem[Grandeit et~al\mbox{.}(2020)]%
        {grandeitUsingBERTQualitative2020}
\bibfield{author}{\bibinfo{person}{Philipp Grandeit}, \bibinfo{person}{Carolyn Haberkern}, \bibinfo{person}{Maximiliane Lang}, \bibinfo{person}{Jens Albrecht}, {and} \bibinfo{person}{Robert Lehmann}.} \bibinfo{year}{2020}\natexlab{}.
\newblock \showarticletitle{Using {{BERT}} for {{Qualitative Content Analysis}} in {{Psychosocial Online Counseling}}}. In \bibinfo{booktitle}{\emph{Proceedings of the {{Fourth Workshop}} on {{Natural Language Processing}} and {{Computational Social Science}}}} (Online, 2020). \bibinfo{publisher}{Association for Computational Linguistics}, \bibinfo{pages}{11--23}.
\newblock
\href{https://doi.org/10.18653/v1/2020.nlpcss-1.2}{doi:\nolinkurl{10.18653/v1/2020.nlpcss-1.2}}


\bibitem[Hamilton et~al\mbox{.}(2023)]%
        {hamiltonExploringUseAI2023}
\bibfield{author}{\bibinfo{person}{Leah Hamilton}, \bibinfo{person}{Desha Elliott}, \bibinfo{person}{Aaron Quick}, \bibinfo{person}{Simone Smith}, {and} \bibinfo{person}{Victoria Choplin}.} \bibinfo{year}{2023}\natexlab{}.
\newblock \showarticletitle{Exploring the {{Use}} of {{AI}} in {{Qualitative Analysis}}: {{A Comparative Study}} of {{Guaranteed Income Data}}}.
\newblock   \bibinfo{volume}{22} (\bibinfo{year}{2023}), \bibinfo{pages}{16094069231201504}.
\newblock
\showISSN{1609-4069, 1609-4069}
\href{https://doi.org/10.1177/16094069231201504}{doi:\nolinkurl{10.1177/16094069231201504}}


\bibitem[He et~al\mbox{.}(2021b)]%
        {heDeBERTaV3ImprovingDeBERTa2021}
\bibfield{author}{\bibinfo{person}{Pengcheng He}, \bibinfo{person}{Jianfeng Gao}, {and} \bibinfo{person}{Weizhu Chen}.} \bibinfo{year}{2021}\natexlab{b}.
\newblock \bibinfo{booktitle}{\emph{{{DeBERTaV3}}: {{Improving DeBERTa}} Using {{ELECTRA-Style Pre-Training}} with {{Gradient-Disentangled Embedding Sharing}}}}.
\newblock
\href{https://doi.org/10.48550/ARXIV.2111.09543}{doi:\nolinkurl{10.48550/ARXIV.2111.09543}}


\bibitem[He et~al\mbox{.}(2020a)]%
        {heDeBERTaDecodingenhancedBERT2020}
\bibfield{author}{\bibinfo{person}{Pengcheng He}, \bibinfo{person}{Xiaodong Liu}, \bibinfo{person}{Jianfeng Gao}, {and} \bibinfo{person}{Weizhu Chen}.} \bibinfo{year}{2020}\natexlab{a}.
\newblock \bibinfo{booktitle}{\emph{{{DeBERTa}}: {{Decoding-enhanced BERT}} with {{Disentangled Attention}}}}.
\newblock
\href{https://doi.org/10.48550/ARXIV.2006.03654}{doi:\nolinkurl{10.48550/ARXIV.2006.03654}}


\bibitem[Holstein and Gubrium(1995)]%
        {holsteinActiveInterview1995}
\bibfield{author}{\bibinfo{person}{James Holstein} {and} \bibinfo{person}{Jaber Gubrium}.} \bibinfo{year}{1995}\natexlab{}.
\newblock \bibinfo{booktitle}{\emph{The {{Active Interview}}}}.
\newblock \bibinfo{publisher}{SAGE Publications, Inc.}, \bibinfo{address}{2455 Teller Road,~Thousand Oaks~California~91320~United States of America}.
\newblock
\showISBNx{978-0-8039-5895-1 978-1-4129-8612-0}
\href{https://doi.org/10.4135/9781412986120}{doi:\nolinkurl{10.4135/9781412986120}}


\bibitem[Holstein et~al\mbox{.}(2019)]%
        {holsteinImprovingFairnessMachine2019}
\bibfield{author}{\bibinfo{person}{Kenneth Holstein}, \bibinfo{person}{Jennifer Wortman~Vaughan}, \bibinfo{person}{Hal Daum{\'e}}, \bibinfo{person}{Miro Dudik}, {and} \bibinfo{person}{Hanna Wallach}.} \bibinfo{year}{2019}\natexlab{}.
\newblock \showarticletitle{Improving {{Fairness}} in {{Machine Learning Systems}}: {{What Do Industry Practitioners Need}}?}. In \bibinfo{booktitle}{\emph{Proceedings of the 2019 {{CHI Conference}} on {{Human Factors}} in {{Computing Systems}}}}. \bibinfo{publisher}{ACM}, \bibinfo{address}{Glasgow Scotland Uk}, \bibinfo{pages}{1--16}.
\newblock
\showISBNx{978-1-4503-5970-2}
\href{https://doi.org/10.1145/3290605.3300830}{doi:\nolinkurl{10.1145/3290605.3300830}}


\bibitem[Homola et~al\mbox{.}(2016)]%
        {homolaMeasureSurveyMode2016}
\bibfield{author}{\bibinfo{person}{Jonathan Homola}, \bibinfo{person}{Natalie Jackson}, {and} \bibinfo{person}{Jeff Gill}.} \bibinfo{year}{2016}\natexlab{}.
\newblock \showarticletitle{A Measure of Survey Mode Differences}.
\newblock   \bibinfo{volume}{44} (\bibinfo{year}{2016}), \bibinfo{pages}{255--274}.
\newblock
\showISSN{0261-3794}
\href{https://doi.org/10.1016/j.electstud.2016.06.010}{doi:\nolinkurl{10.1016/j.electstud.2016.06.010}}


\bibitem[Koutsoumpis et~al\mbox{.}(2024)]%
        {koutsoumpis2024}
\bibfield{author}{\bibinfo{person}{Alexandros Koutsoumpis}, \bibinfo{person}{Shadi Ghassemi}, \bibinfo{person}{Janneke~K Oostrom}, {and} \bibinfo{person}{Daan Holtrop}.} \bibinfo{year}{2024}\natexlab{}.
\newblock \showarticletitle{AI-Assisted Video Interviews: Psychometric Analysis of Asynchronous Video Interviews for Personality and Interview Performance Evaluation using Machine Learning}.
\newblock \bibinfo{journal}{\emph{Computers in Human Behavior}}  \bibinfo{volume}{152} (\bibinfo{year}{2024}), \bibinfo{pages}{108045}.
\newblock


\bibitem[Krishnan et~al\mbox{.}(2023)]%
        {krishnanArtificialIntelligenceClinical2023}
\bibfield{author}{\bibinfo{person}{Gokul Krishnan}, \bibinfo{person}{Shiana Singh}, \bibinfo{person}{Monika Pathania}, \bibinfo{person}{Siddharth Gosavi}, \bibinfo{person}{Shuchi Abhishek}, \bibinfo{person}{Ashwin Parchani}, {and} \bibinfo{person}{Minakshi Dhar}.} \bibinfo{year}{2023}\natexlab{}.
\newblock \showarticletitle{Artificial Intelligence in Clinical Medicine: Catalyzing a Sustainable Global Healthcare Paradigm}.
\newblock \bibinfo{journal}{\emph{Frontiers in Artificial Intelligence}}  \bibinfo{volume}{6} (\bibinfo{date}{Aug.} \bibinfo{year}{2023}), \bibinfo{pages}{1227091}.
\newblock
\showISSN{2624-8212}
\href{https://doi.org/10.3389/frai.2023.1227091}{doi:\nolinkurl{10.3389/frai.2023.1227091}}


\bibitem[Kvale and Brinkmann(2021)]%
        {kvale2021}
\bibfield{author}{\bibinfo{person}{Steinar Kvale} {and} \bibinfo{person}{Svend Brinkmann}.} \bibinfo{year}{2021}\natexlab{}.
\newblock \bibinfo{booktitle}{\emph{Doing Interviews} (\bibinfo{edition}{2nd} ed.)}.
\newblock \bibinfo{publisher}{SAGE Publications}.
\newblock


\bibitem[Lacan et~al\mbox{.}(2006)]%
        {lacanEcritsFirstComplete2006}
\bibfield{author}{\bibinfo{person}{Jacques Lacan}, \bibinfo{person}{Bruce Fink}, {and} \bibinfo{person}{Jacques Lacan}.} \bibinfo{year}{2006}\natexlab{}.
\newblock \bibinfo{booktitle}{\emph{Ecrits: {{The}} First Complete Edition in {{English}}}}.
\newblock \bibinfo{publisher}{Norton}.
\newblock
\showISBNx{978-0-393-32925-4 978-0-393-06115-4}


\bibitem[Li et~al\mbox{.}(2024)]%
        {liComparingGPT4Human2024}
\bibfield{author}{\bibinfo{person}{Kevin~Danis Li}, \bibinfo{person}{Adrian~M Fernandez}, \bibinfo{person}{Rachel Schwartz}, \bibinfo{person}{Natalie Rios}, \bibinfo{person}{Marvin~Nathaniel Carlisle}, \bibinfo{person}{Gregory~M Amend}, \bibinfo{person}{Hiren~V Patel}, {and} \bibinfo{person}{Benjamin~N Breyer}.} \bibinfo{year}{2024}\natexlab{}.
\newblock \showarticletitle{Comparing {{GPT-4}} and {{Human Researchers}} in {{Health Care Data Analysis}}: {{Qualitative Description Study}}}.
\newblock   \bibinfo{volume}{26} (\bibinfo{year}{2024}), \bibinfo{pages}{e56500}.
\newblock
\showISSN{1438-8871}
\href{https://doi.org/10.2196/56500}{doi:\nolinkurl{10.2196/56500}}


\bibitem[Luger and Sellen(2016)]%
        {lugerHavingReallyBad2016}
\bibfield{author}{\bibinfo{person}{Ewa Luger} {and} \bibinfo{person}{Abigail Sellen}.} \bibinfo{year}{2016}\natexlab{}.
\newblock \showarticletitle{"{{Like Having}} a {{Really Bad PA}}": {{The Gulf}} between {{User Expectation}} and {{Experience}} of {{Conversational Agents}}}. In \bibinfo{booktitle}{\emph{Proceedings of the 2016 {{CHI Conference}} on {{Human Factors}} in {{Computing Systems}}}} (San Jose California USA, 2016-05-07). \bibinfo{publisher}{ACM}, \bibinfo{pages}{5286--5297}.
\newblock
\showISBNx{978-1-4503-3362-7}
\href{https://doi.org/10.1145/2858036.2858288}{doi:\nolinkurl{10.1145/2858036.2858288}}


\bibitem[Luo et~al\mbox{.}(2025)]%
        {luoWhyMightAIenabled2025}
\bibfield{author}{\bibinfo{person}{Wenhao Luo}, \bibinfo{person}{Yuelin Zhang}, {and} \bibinfo{person}{Maona Mu}.} \bibinfo{year}{2025}\natexlab{}.
\newblock \showarticletitle{Why Might {{AI-enabled}} Interviews Reduce Candidates’ Job Application Intention? {{The}} Role of Procedural Justice and Organizational Attractiveness}.
\newblock  \bibinfo{volume}{12}, \bibinfo{number}{1} (\bibinfo{year}{2025}), \bibinfo{pages}{1278}.
\newblock
\showISSN{2662-9992}
\href{https://doi.org/10.1057/s41599-025-05607-z}{doi:\nolinkurl{10.1057/s41599-025-05607-z}}


\bibitem[Mostafavi et~al\mbox{.}(2025)]%
        {mostafaviContextualEmbeddingsSociological2025}
\bibfield{author}{\bibinfo{person}{Moeen Mostafavi}, \bibinfo{person}{Michael~D. Porter}, {and} \bibinfo{person}{Dawn~T. Robinson}.} \bibinfo{year}{2025}\natexlab{}.
\newblock \showarticletitle{Contextual {{Embeddings}} in {{Sociological Research}}: {{Expanding}} the {{Analysis}} of {{Sentiment}} and {{Social Dynamics}}}.
\newblock  \bibinfo{volume}{55}, \bibinfo{number}{1} (\bibinfo{year}{2025}), \bibinfo{pages}{25--58}.
\newblock
\showISSN{0081-1750, 1467-9531}
\href{https://doi.org/10.1177/00811750241260729}{doi:\nolinkurl{10.1177/00811750241260729}}


\bibitem[Nass and Moon(2000)]%
        {nassMachinesMindlessnessSocial2000}
\bibfield{author}{\bibinfo{person}{Clifford Nass} {and} \bibinfo{person}{Youngme Moon}.} \bibinfo{year}{2000}\natexlab{}.
\newblock \showarticletitle{Machines and {{Mindlessness}}: {{Social Responses}} to {{Computers}}}.
\newblock  \bibinfo{volume}{56}, \bibinfo{number}{1} (\bibinfo{year}{2000}), \bibinfo{pages}{81--103}.
\newblock
\showISSN{0022-4537, 1540-4560}
\href{https://doi.org/10.1111/0022-4537.00153}{doi:\nolinkurl{10.1111/0022-4537.00153}}


\bibitem[OpenAI et~al\mbox{.}(2024)]%
        {openaiGPT4TechnicalReport2024}
\bibfield{author}{\bibinfo{person}{OpenAI}, \bibinfo{person}{Josh Achiam}, \bibinfo{person}{Steven Adler}, \bibinfo{person}{Sandhini Agarwal}, \bibinfo{person}{Lama Ahmad}, \bibinfo{person}{Ilge Akkaya}, \bibinfo{person}{Florencia~Leoni Aleman}, \bibinfo{person}{Diogo Almeida}, \bibinfo{person}{Janko Altenschmidt}, \bibinfo{person}{Sam Altman}, \bibinfo{person}{Shyamal Anadkat}, \bibinfo{person}{Red Avila}, \bibinfo{person}{Igor Babuschkin}, \bibinfo{person}{Suchir Balaji}, \bibinfo{person}{Valerie Balcom}, \bibinfo{person}{Paul Baltescu}, \bibinfo{person}{Haiming Bao}, \bibinfo{person}{Mohammad Bavarian}, \bibinfo{person}{Jeff Belgum}, \bibinfo{person}{Irwan Bello}, \bibinfo{person}{Jake Berdine}, \bibinfo{person}{Gabriel {Bernadett-Shapiro}}, \bibinfo{person}{Christopher Berner}, \bibinfo{person}{Lenny Bogdonoff}, \bibinfo{person}{Oleg Boiko}, \bibinfo{person}{Madelaine Boyd}, \bibinfo{person}{Anna-Luisa Brakman}, \bibinfo{person}{Greg Brockman}, \bibinfo{person}{Tim Brooks}, \bibinfo{person}{Miles Brundage}, \bibinfo{person}{Kevin Button}, \bibinfo{person}{Trevor Cai}, \bibinfo{person}{Rosie Campbell}, \bibinfo{person}{Andrew Cann}, \bibinfo{person}{Brittany Carey}, \bibinfo{person}{Chelsea Carlson}, \bibinfo{person}{Rory Carmichael}, \bibinfo{person}{Brooke Chan}, \bibinfo{person}{Che Chang}, \bibinfo{person}{Fotis Chantzis}, \bibinfo{person}{Derek Chen}, \bibinfo{person}{Sully Chen}, \bibinfo{person}{Ruby Chen}, \bibinfo{person}{Jason Chen}, \bibinfo{person}{Mark Chen}, \bibinfo{person}{Ben Chess}, \bibinfo{person}{Chester Cho}, \bibinfo{person}{Casey Chu}, \bibinfo{person}{Hyung~Won Chung}, \bibinfo{person}{Dave Cummings}, \bibinfo{person}{Jeremiah Currier}, \bibinfo{person}{Yunxing Dai}, \bibinfo{person}{Cory Decareaux}, \bibinfo{person}{Thomas Degry}, \bibinfo{person}{Noah Deutsch}, \bibinfo{person}{Damien Deville}, \bibinfo{person}{Arka Dhar}, \bibinfo{person}{David Dohan}, \bibinfo{person}{Steve Dowling}, \bibinfo{person}{Sheila Dunning}, \bibinfo{person}{Adrien Ecoffet}, \bibinfo{person}{Atty Eleti}, \bibinfo{person}{Tyna Eloundou}, \bibinfo{person}{David Farhi}, \bibinfo{person}{Liam Fedus}, \bibinfo{person}{Niko Felix}, \bibinfo{person}{Sim{\'o}n~Posada Fishman}, \bibinfo{person}{Juston Forte}, \bibinfo{person}{Isabella Fulford}, \bibinfo{person}{Leo Gao}, \bibinfo{person}{Elie Georges}, \bibinfo{person}{Christian Gibson}, \bibinfo{person}{Vik Goel}, \bibinfo{person}{Tarun Gogineni}, \bibinfo{person}{Gabriel Goh}, \bibinfo{person}{Rapha {Gontijo-Lopes}}, \bibinfo{person}{Jonathan Gordon}, \bibinfo{person}{Morgan Grafstein}, \bibinfo{person}{Scott Gray}, \bibinfo{person}{Ryan Greene}, \bibinfo{person}{Joshua Gross}, \bibinfo{person}{Shixiang~Shane Gu}, \bibinfo{person}{Yufei Guo}, \bibinfo{person}{Chris Hallacy}, \bibinfo{person}{Jesse Han}, \bibinfo{person}{Jeff Harris}, \bibinfo{person}{Yuchen He}, \bibinfo{person}{Mike Heaton}, \bibinfo{person}{Johannes Heidecke}, \bibinfo{person}{Chris Hesse}, \bibinfo{person}{Alan Hickey}, \bibinfo{person}{Wade Hickey}, \bibinfo{person}{Peter Hoeschele}, \bibinfo{person}{Brandon Houghton}, \bibinfo{person}{Kenny Hsu}, \bibinfo{person}{Shengli Hu}, \bibinfo{person}{Xin Hu}, \bibinfo{person}{Joost Huizinga}, \bibinfo{person}{Shantanu Jain}, \bibinfo{person}{Shawn Jain}, \bibinfo{person}{Joanne Jang}, \bibinfo{person}{Angela Jiang}, \bibinfo{person}{Roger Jiang}, \bibinfo{person}{Haozhun Jin}, \bibinfo{person}{Denny Jin}, \bibinfo{person}{Shino Jomoto}, \bibinfo{person}{Billie Jonn}, \bibinfo{person}{Heewoo Jun}, \bibinfo{person}{Tomer Kaftan}, \bibinfo{person}{{\L}ukasz Kaiser}, \bibinfo{person}{Ali Kamali}, \bibinfo{person}{Ingmar Kanitscheider}, \bibinfo{person}{Nitish~Shirish Keskar}, \bibinfo{person}{Tabarak Khan}, \bibinfo{person}{Logan Kilpatrick}, \bibinfo{person}{Jong~Wook Kim}, \bibinfo{person}{Christina Kim}, \bibinfo{person}{Yongjik Kim}, \bibinfo{person}{Jan~Hendrik Kirchner}, \bibinfo{person}{Jamie Kiros}, \bibinfo{person}{Matt Knight}, \bibinfo{person}{Daniel Kokotajlo}, \bibinfo{person}{{\L}ukasz Kondraciuk}, \bibinfo{person}{Andrew Kondrich}, \bibinfo{person}{Aris Konstantinidis}, \bibinfo{person}{Kyle Kosic}, \bibinfo{person}{Gretchen Krueger}, \bibinfo{person}{Vishal Kuo}, \bibinfo{person}{Michael Lampe}, \bibinfo{person}{Ikai Lan}, \bibinfo{person}{Teddy Lee}, \bibinfo{person}{Jan Leike}, \bibinfo{person}{Jade Leung}, \bibinfo{person}{Daniel Levy}, \bibinfo{person}{Chak~Ming Li}, \bibinfo{person}{Rachel Lim}, \bibinfo{person}{Molly Lin}, \bibinfo{person}{Stephanie Lin}, \bibinfo{person}{Mateusz Litwin}, \bibinfo{person}{Theresa Lopez}, \bibinfo{person}{Ryan Lowe}, \bibinfo{person}{Patricia Lue}, \bibinfo{person}{Anna Makanju}, \bibinfo{person}{Kim Malfacini}, \bibinfo{person}{Sam Manning}, \bibinfo{person}{Todor Markov}, \bibinfo{person}{Yaniv Markovski}, \bibinfo{person}{Bianca Martin}, \bibinfo{person}{Katie Mayer}, \bibinfo{person}{Andrew Mayne}, \bibinfo{person}{Bob McGrew}, \bibinfo{person}{Scott~Mayer McKinney}, \bibinfo{person}{Christine McLeavey}, \bibinfo{person}{Paul McMillan}, \bibinfo{person}{Jake McNeil}, \bibinfo{person}{David Medina}, \bibinfo{person}{Aalok Mehta}, \bibinfo{person}{Jacob Menick}, \bibinfo{person}{Luke Metz}, \bibinfo{person}{Andrey Mishchenko}, \bibinfo{person}{Pamela Mishkin}, \bibinfo{person}{Vinnie Monaco}, \bibinfo{person}{Evan Morikawa}, \bibinfo{person}{Daniel Mossing}, \bibinfo{person}{Tong Mu}, \bibinfo{person}{Mira Murati}, \bibinfo{person}{Oleg Murk}, \bibinfo{person}{David M{\'e}ly}, \bibinfo{person}{Ashvin Nair}, \bibinfo{person}{Reiichiro Nakano}, \bibinfo{person}{Rajeev Nayak}, \bibinfo{person}{Arvind Neelakantan}, \bibinfo{person}{Richard Ngo}, \bibinfo{person}{Hyeonwoo Noh}, \bibinfo{person}{Long Ouyang}, \bibinfo{person}{Cullen O'Keefe}, \bibinfo{person}{Jakub Pachocki}, \bibinfo{person}{Alex Paino}, \bibinfo{person}{Joe Palermo}, \bibinfo{person}{Ashley Pantuliano}, \bibinfo{person}{Giambattista Parascandolo}, \bibinfo{person}{Joel Parish}, \bibinfo{person}{Emy Parparita}, \bibinfo{person}{Alex Passos}, \bibinfo{person}{Mikhail Pavlov}, \bibinfo{person}{Andrew Peng}, \bibinfo{person}{Adam Perelman}, \bibinfo{person}{Filipe de Avila~Belbute Peres}, \bibinfo{person}{Michael Petrov}, \bibinfo{person}{Henrique Ponde de~Oliveira Pinto}, \bibinfo{person}{Michael}, \bibinfo{person}{Pokorny}, \bibinfo{person}{Michelle Pokrass}, \bibinfo{person}{Vitchyr~H. Pong}, \bibinfo{person}{Tolly Powell}, \bibinfo{person}{Alethea Power}, \bibinfo{person}{Boris Power}, \bibinfo{person}{Elizabeth Proehl}, \bibinfo{person}{Raul Puri}, \bibinfo{person}{Alec Radford}, \bibinfo{person}{Jack Rae}, \bibinfo{person}{Aditya Ramesh}, \bibinfo{person}{Cameron Raymond}, \bibinfo{person}{Francis Real}, \bibinfo{person}{Kendra Rimbach}, \bibinfo{person}{Carl Ross}, \bibinfo{person}{Bob Rotsted}, \bibinfo{person}{Henri Roussez}, \bibinfo{person}{Nick Ryder}, \bibinfo{person}{Mario Saltarelli}, \bibinfo{person}{Ted Sanders}, \bibinfo{person}{Shibani Santurkar}, \bibinfo{person}{Girish Sastry}, \bibinfo{person}{Heather Schmidt}, \bibinfo{person}{David Schnurr}, \bibinfo{person}{John Schulman}, \bibinfo{person}{Daniel Selsam}, \bibinfo{person}{Kyla Sheppard}, \bibinfo{person}{Toki Sherbakov}, \bibinfo{person}{Jessica Shieh}, \bibinfo{person}{Sarah Shoker}, \bibinfo{person}{Pranav Shyam}, \bibinfo{person}{Szymon Sidor}, \bibinfo{person}{Eric Sigler}, \bibinfo{person}{Maddie Simens}, \bibinfo{person}{Jordan Sitkin}, \bibinfo{person}{Katarina Slama}, \bibinfo{person}{Ian Sohl}, \bibinfo{person}{Benjamin Sokolowsky}, \bibinfo{person}{Yang Song}, \bibinfo{person}{Natalie Staudacher}, \bibinfo{person}{Felipe~Petroski Such}, \bibinfo{person}{Natalie Summers}, \bibinfo{person}{Ilya Sutskever}, \bibinfo{person}{Jie Tang}, \bibinfo{person}{Nikolas Tezak}, \bibinfo{person}{Madeleine~B. Thompson}, \bibinfo{person}{Phil Tillet}, \bibinfo{person}{Amin Tootoonchian}, \bibinfo{person}{Elizabeth Tseng}, \bibinfo{person}{Preston Tuggle}, \bibinfo{person}{Nick Turley}, \bibinfo{person}{Jerry Tworek}, \bibinfo{person}{Juan Felipe~Cer{\'o}n Uribe}, \bibinfo{person}{Andrea Vallone}, \bibinfo{person}{Arun Vijayvergiya}, \bibinfo{person}{Chelsea Voss}, \bibinfo{person}{Carroll Wainwright}, \bibinfo{person}{Justin~Jay Wang}, \bibinfo{person}{Alvin Wang}, \bibinfo{person}{Ben Wang}, \bibinfo{person}{Jonathan Ward}, \bibinfo{person}{Jason Wei}, \bibinfo{person}{C.~J. Weinmann}, \bibinfo{person}{Akila Welihinda}, \bibinfo{person}{Peter Welinder}, \bibinfo{person}{Jiayi Weng}, \bibinfo{person}{Lilian Weng}, \bibinfo{person}{Matt Wiethoff}, \bibinfo{person}{Dave Willner}, \bibinfo{person}{Clemens Winter}, \bibinfo{person}{Samuel Wolrich}, \bibinfo{person}{Hannah Wong}, \bibinfo{person}{Lauren Workman}, \bibinfo{person}{Sherwin Wu}, \bibinfo{person}{Jeff Wu}, \bibinfo{person}{Michael Wu}, \bibinfo{person}{Kai Xiao}, \bibinfo{person}{Tao Xu}, \bibinfo{person}{Sarah Yoo}, \bibinfo{person}{Kevin Yu}, \bibinfo{person}{Qiming Yuan}, \bibinfo{person}{Wojciech Zaremba}, \bibinfo{person}{Rowan Zellers}, \bibinfo{person}{Chong Zhang}, \bibinfo{person}{Marvin Zhang}, \bibinfo{person}{Shengjia Zhao}, \bibinfo{person}{Tianhao Zheng}, \bibinfo{person}{Juntang Zhuang}, \bibinfo{person}{William Zhuk}, {and} \bibinfo{person}{Barret Zoph}.} \bibinfo{year}{2024}\natexlab{}.
\newblock \bibinfo{title}{{{GPT-4 Technical Report}}}.
\newblock
\showeprint[arxiv]{2303.08774}~[cs]
\href{https://doi.org/10.48550/arXiv.2303.08774}{doi:\nolinkurl{10.48550/arXiv.2303.08774}}


\bibitem[Pichard and Roster(2020)]%
        {pichardRoster2020}
\bibfield{author}{\bibinfo{person}{Marie-Claude Pichard} {and} \bibinfo{person}{Catherine~A Roster}.} \bibinfo{year}{2020}\natexlab{}.
\newblock \showarticletitle{The Dark Side of Digital Disclosure: An Investigation of the Unintended Consequences of Digital Self-Disclosure for Consumers}.
\newblock \bibinfo{journal}{\emph{Journal of Business Research}}  \bibinfo{volume}{120} (\bibinfo{year}{2020}), \bibinfo{pages}{169--178}.
\newblock


\bibitem[Pickard and Roster(2020)]%
        {pickardUsingComputerAutomated2020}
\bibfield{author}{\bibinfo{person}{Matthew~D. Pickard} {and} \bibinfo{person}{Catherine~A. Roster}.} \bibinfo{year}{2020}\natexlab{}.
\newblock \showarticletitle{Using Computer Automated Systems to Conduct Personal Interviews: {{Does}} the Mere Presence of a Human Face Inhibit Disclosure?}
\newblock \bibinfo{journal}{\emph{Computers in Human Behavior}}  \bibinfo{volume}{105} (\bibinfo{date}{April} \bibinfo{year}{2020}), \bibinfo{pages}{106197}.
\newblock
\showISSN{07475632}
\href{https://doi.org/10.1016/j.chb.2019.106197}{doi:\nolinkurl{10.1016/j.chb.2019.106197}}


\bibitem[Prescott et~al\mbox{.}(2024)]%
        {prescottComparingEfficacyEfficiency2024}
\bibfield{author}{\bibinfo{person}{Maximo~R Prescott}, \bibinfo{person}{Samantha Yeager}, \bibinfo{person}{Lillian Ham}, \bibinfo{person}{Carlos~D Rivera~Saldana}, \bibinfo{person}{Vanessa Serrano}, \bibinfo{person}{Joey Narez}, \bibinfo{person}{Dafna Paltin}, \bibinfo{person}{Jorge Delgado}, \bibinfo{person}{David~J Moore}, {and} \bibinfo{person}{Jessica Montoya}.} \bibinfo{year}{2024}\natexlab{}.
\newblock \showarticletitle{Comparing the {{Efficacy}} and {{Efficiency}} of {{Human}} and {{Generative AI}}: {{Qualitative Thematic Analyses}}}.
\newblock   \bibinfo{volume}{3} (\bibinfo{year}{2024}), \bibinfo{pages}{e54482}.
\newblock
\showISSN{2817-1705}
\href{https://doi.org/10.2196/54482}{doi:\nolinkurl{10.2196/54482}}


\bibitem[Sakaguchi et~al\mbox{.}(2025)]%
        {sakaguchiEvaluatingChatGPTQualitative2025}
\bibfield{author}{\bibinfo{person}{Kota Sakaguchi}, \bibinfo{person}{Reiko Sakama}, {and} \bibinfo{person}{Takashi Watari}.} \bibinfo{year}{2025}\natexlab{}.
\newblock \showarticletitle{Evaluating {{ChatGPT}} in {{Qualitative Thematic Analysis With Human Researchers}} in the {{Japanese Clinical Context}} and {{Its Cultural Interpretation Challenges}}: {{Comparative Qualitative Study}}}.
\newblock   \bibinfo{volume}{27} (\bibinfo{year}{2025}), \bibinfo{pages}{e71521}.
\newblock
\showISSN{1438-8871}
\href{https://doi.org/10.2196/71521}{doi:\nolinkurl{10.2196/71521}}


\bibitem[Sapci and Sapci(2020)]%
        {sapciArtificialIntelligenceEducation2020}
\bibfield{author}{\bibinfo{person}{A~Hasan Sapci} {and} \bibinfo{person}{H~Aylin Sapci}.} \bibinfo{year}{2020}\natexlab{}.
\newblock \showarticletitle{Artificial {{Intelligence Education}} and {{Tools}} for {{Medical}} and {{Health Informatics Students}}: {{Systematic Review}}}.
\newblock \bibinfo{journal}{\emph{JMIR Medical Education}} \bibinfo{volume}{6}, \bibinfo{number}{1} (\bibinfo{date}{June} \bibinfo{year}{2020}), \bibinfo{pages}{e19285}.
\newblock
\showISSN{2369-3762}
\href{https://doi.org/10.2196/19285}{doi:\nolinkurl{10.2196/19285}}


\bibitem[Searle(1980)]%
        {searleMindsBrainsPrograms1980}
\bibfield{author}{\bibinfo{person}{John~R. Searle}.} \bibinfo{year}{1980}\natexlab{}.
\newblock \showarticletitle{Minds, Brains, and Programs}.
\newblock  \bibinfo{volume}{3}, \bibinfo{number}{3} (\bibinfo{year}{1980}), \bibinfo{pages}{417--424}.
\newblock
\showISSN{0140-525X, 1469-1825}
\href{https://doi.org/10.1017/S0140525X00005756}{doi:\nolinkurl{10.1017/S0140525X00005756}}


\bibitem[Seidman(2019)]%
        {seidman2019}
\bibfield{author}{\bibinfo{person}{Irving Seidman}.} \bibinfo{year}{2019}\natexlab{}.
\newblock \bibinfo{booktitle}{\emph{Interviewing as Qualitative Research: A Guide for Researchers in Education and the Social Sciences} (\bibinfo{edition}{5th} ed.)}.
\newblock \bibinfo{publisher}{Teachers College Press}.
\newblock


\bibitem[Shapka et~al\mbox{.}(2016)]%
        {shapkaOnlineInpersonInterviews2016}
\bibfield{author}{\bibinfo{person}{Jennifer~D. Shapka}, \bibinfo{person}{Jose~F. Domene}, \bibinfo{person}{Shereen Khan}, {and} \bibinfo{person}{Leigh~Mijin Yang}.} \bibinfo{year}{2016}\natexlab{}.
\newblock \showarticletitle{Online versus In-Person Interviews with Adolescents: {{An}} Exploration of Data Equivalence}.
\newblock \bibinfo{journal}{\emph{Computers in Human Behavior}}  \bibinfo{volume}{58} (\bibinfo{date}{May} \bibinfo{year}{2016}), \bibinfo{pages}{361--367}.
\newblock
\showISSN{07475632}
\href{https://doi.org/10.1016/j.chb.2016.01.016}{doi:\nolinkurl{10.1016/j.chb.2016.01.016}}


\bibitem[Shuster et~al\mbox{.}(2021)]%
        {shusterRetrievalAugmentationReduces2021}
\bibfield{author}{\bibinfo{person}{Kurt Shuster}, \bibinfo{person}{Spencer Poff}, \bibinfo{person}{Moya Chen}, \bibinfo{person}{Douwe Kiela}, {and} \bibinfo{person}{Jason Weston}.} \bibinfo{year}{2021}\natexlab{}.
\newblock \bibinfo{title}{Retrieval {{Augmentation Reduces Hallucination}} in {{Conversation}}}.
\newblock
\showeprint[arxiv]{2104.07567}~[cs]
\href{https://doi.org/10.48550/arXiv.2104.07567}{doi:\nolinkurl{10.48550/arXiv.2104.07567}}


\bibitem[Stokols(2006)]%
        {stokolsScienceTransdisciplinaryAction2006}
\bibfield{author}{\bibinfo{person}{Daniel Stokols}.} \bibinfo{year}{2006}\natexlab{}.
\newblock \showarticletitle{Toward a {{Science}} of {{Transdisciplinary Action Research}}}.
\newblock  \bibinfo{volume}{38}, \bibinfo{number}{1--2} (\bibinfo{year}{2006}), \bibinfo{pages}{79--93}.
\newblock
\showISSN{00910562}
\href{https://doi.org/10.1007/s10464-006-9060-5}{doi:\nolinkurl{10.1007/s10464-006-9060-5}}


\bibitem[Suen and Hung(2023)]%
        {suenHung2023}
\bibfield{author}{\bibinfo{person}{Hung-Yue Suen} {and} \bibinfo{person}{Kuo-En Hung}.} \bibinfo{year}{2023}\natexlab{}.
\newblock \showarticletitle{Building Trust in Automatic Video Interview Assessment: The Role of Explainability, Tangibility and Transparency}.
\newblock \bibinfo{journal}{\emph{Computers in Human Behavior}}  \bibinfo{volume}{143} (\bibinfo{year}{2023}), \bibinfo{pages}{107713}.
\newblock


\bibitem[Timmons et~al\mbox{.}(2023)]%
        {timmonsCallActionAssessing2023}
\bibfield{author}{\bibinfo{person}{Adela~C. Timmons}, \bibinfo{person}{Jacqueline~B. Duong}, \bibinfo{person}{Natalia Simo~Fiallo}, \bibinfo{person}{Theodore Lee}, \bibinfo{person}{Huong Phuc~Quynh Vo}, \bibinfo{person}{Matthew~W. Ahle}, \bibinfo{person}{Jonathan~S. Comer}, \bibinfo{person}{LaPrincess~C. Brewer}, \bibinfo{person}{Stacy~L. Frazier}, {and} \bibinfo{person}{Theodora Chaspari}.} \bibinfo{year}{2023}\natexlab{}.
\newblock \showarticletitle{A {{Call}} to {{Action}} on {{Assessing}} and {{Mitigating Bias}} in {{Artificial Intelligence Applications}} for {{Mental Health}}}.
\newblock  \bibinfo{volume}{18}, \bibinfo{number}{5} (\bibinfo{year}{2023}), \bibinfo{pages}{1062--1096}.
\newblock
\showISSN{1745-6916, 1745-6924}
\href{https://doi.org/10.1177/17456916221134490}{doi:\nolinkurl{10.1177/17456916221134490}}


\bibitem[Turkle(2011)]%
        {turkleAloneTogetherWhy2011}
\bibfield{author}{\bibinfo{person}{Sherry Turkle}.} \bibinfo{year}{2011}\natexlab{}.
\newblock \bibinfo{booktitle}{\emph{Alone Together: Why We Expect More from Technology and Less from Each Other}}.
\newblock \bibinfo{publisher}{Basic Books}.
\newblock
\showISBNx{978-0-465-01021-9 978-0-465-02234-2}


\bibitem[Vaswani et~al\mbox{.}(2017)]%
        {vaswani2017attention}
\bibfield{author}{\bibinfo{person}{Ashish Vaswani}, \bibinfo{person}{Noam Shazeer}, \bibinfo{person}{Niki Parmar}, \bibinfo{person}{Jakob Uszkoreit}, \bibinfo{person}{Llion Jones}, \bibinfo{person}{Aidan~N Gomez}, \bibinfo{person}{\L~ukasz Kaiser}, {and} \bibinfo{person}{Illia Polosukhin}.} \bibinfo{year}{2017}\natexlab{}.
\newblock \showarticletitle{Attention is All You Need}. In \bibinfo{booktitle}{\emph{Advances in Neural Information Processing Systems}}, Vol.~\bibinfo{volume}{30}. \bibinfo{pages}{5998--6008}.
\newblock


\bibitem[Wankmüller(2021)]%
        {wankmullerIntroductionNeuralTransfer2021}
\bibfield{author}{\bibinfo{person}{Sandra Wankmüller}.} \bibinfo{year}{2021}\natexlab{}.
\newblock \bibinfo{booktitle}{\emph{Introduction to {{Neural Transfer Learning}} with {{Transformers}} for {{Social Science Text Analysis}}}}.
\newblock
\href{https://doi.org/10.48550/ARXIV.2102.02111}{doi:\nolinkurl{10.48550/ARXIV.2102.02111}}


\bibitem[Wuttke et~al\mbox{.}(2024)]%
        {wuttkeAIConversationalInterviewing2024}
\bibfield{author}{\bibinfo{person}{Alexander Wuttke}, \bibinfo{person}{Matthias A{\ss}enmacher}, \bibinfo{person}{Christopher Klamm}, \bibinfo{person}{Max~M. Lang}, \bibinfo{person}{Quirin W{\"u}rschinger}, {and} \bibinfo{person}{Frauke Kreuter}.} \bibinfo{year}{2024}\natexlab{}.
\newblock \bibinfo{title}{{{AI Conversational Interviewing}}: {{Transforming Surveys}} with {{LLMs}} as {{Adaptive Interviewers}}}.
\newblock
\showeprint[arxiv]{2410.01824}~[cs]
\href{https://doi.org/10.48550/arXiv.2410.01824}{doi:\nolinkurl{10.48550/arXiv.2410.01824}}


\bibitem[Zhu et~al\mbox{.}(2021)]%
        {zhuMediaSumLargescaleMedia2021}
\bibfield{author}{\bibinfo{person}{Chenguang Zhu}, \bibinfo{person}{Yang Liu}, \bibinfo{person}{Jie Mei}, {and} \bibinfo{person}{Michael Zeng}.} \bibinfo{year}{2021}\natexlab{}.
\newblock \showarticletitle{{{MediaSum}}: {{A Large-scale Media Interview Dataset}} for {{Dialogue Summarization}}}. In \bibinfo{booktitle}{\emph{Proceedings of the 2021 {{Conference}} of the {{North American Chapter}} of the {{Association}} for {{Computational Linguistics}}: {{Human Language Technologies}}}} (Online, 2021). \bibinfo{publisher}{Association for Computational Linguistics}, \bibinfo{pages}{5927--5934}.
\newblock
\href{https://doi.org/10.18653/v1/2021.naacl-main.474}{doi:\nolinkurl{10.18653/v1/2021.naacl-main.474}}


\bibitem[Žižek(2002)]%
        {zizekSublimeObjectIdeology2002}
\bibfield{author}{\bibinfo{person}{Slavoj Žižek}.} \bibinfo{year}{2002}\natexlab{}.
\newblock \bibinfo{booktitle}{\emph{The Sublime Object of Ideology} (\bibinfo{edition}{9. impression} ed.)}.
\newblock \bibinfo{publisher}{Verso}.
\newblock
\showISBNx{978-0-86091-971-1 978-0-86091-256-9}


\end{thebibliography}

\pagebreak

\appendix

\section{Usability Test Interview Outline}
\label{sec:usability_test_outline}
\begin{lstlisting}
    Interview Outline:
    "briefly tell yourself
    has somebody interviewed you, what do you feel?
    If yes, is it a face-to-face interview, online interview, or telephone interview?
    What do you think of this interview app? 
    What do you think are the weaknesses and strengths?
    Is there anything you'd like to add?
    Ask no more than 7 questions in total. 
    Afterwards, thank the participant and tell them the code for prolific is CIxxxxx"
\end{lstlisting}   
\newpage
\section{Participant Demographics and Characteristics for Usability Study}
\label{sec:usability_demographics}
\begin{table}[h]
    \caption{Participant Demographics and Characteristics for Usability Study}
    \label{tab:usability_demographics}
    \small
    \begin{tabular}{@{}p{4cm}|p{8cm}@{}}
    \toprule
    \textbf{Characteristic} & \textbf{Details} \\
    \midrule
    Total Participants & 20 completed interviews (41 initial clicks) \\
    Age Range & 18-70 years (median: early 30s) \\
    Gender Distribution & 14 female, 12 male, 3 undisclosed \\
    Educational Level & High school/A-Levels to Master's degrees; Bachelor's most common \\
    Geographic Location & Predominantly United Kingdom \\
    Occupational Profile & Students, HR advisors, financial analysts, teachers, healthcare workers, administrative staff, homemakers, retired individuals \\
    Prior Interview Experience & Majority experienced in job-seeking contexts across face-to-face, online, and telephone formats \\
    Technical Literacy & High (recruited via Prolific platform) \\
    \bottomrule
    \end{tabular}
\end{table}
\newpage
\section{Participant Demographics and Research Background for Study 3}
\label{sec:participant_details_study_3}
\begin{table}[h]
    \caption{Participant Demographics and Research Background for Study 3. Ten academic researchers from diverse disciplines (philosophy, microbiology, computer science, economics, atmospheric science, etc.) with varying experience levels (PhD students to professors) participated in both human-led and AI-led interviews, with AI effectiveness ratings ranging from 3.0 to 5.0.}
    \label{tab:participant_details}
    \small
    \begin{tabular}{@{}p{0.6cm}p{2.8cm}p{2.2cm}p{1.8cm}p{1.2cm}p{4.8cm}@{}}
    \toprule
    \textbf{ID} & \textbf{Research Field} & \textbf{Experience Level} & \textbf{Interview Type} & \textbf{AI Rating} & \textbf{Key Research Focus} \\
    \midrule
    P1 & Philosophy / Marxist Theory & 0.5 years (Lecturer) & Human-led & 3.5/5 & Dialectical materialism, social theory \\
    P2 & Microbiology & PhD 3rd year & Human-led & 3.0/5 & Bacterial genetics, antimicrobial resistance \\
    P3 & Computer Science (Vehicular Networks) & PhD 3rd year & Human-led & 4.0/5 & Vehicle-to-vehicle communication, network protocols \\
    P4 & Economics (Economic History) & PhD 4th year & Human-led & 3.5/5 & Industrial development, economic transformation \\
    P5 & Innovation \& Strategy (Human-AI) & PhD 5th year & Human-led & 4.5/5 & Human-AI collaboration, organizational innovation \\
    P6 & Atmospheric Science & PhD 2nd year & AI-led & 3.5/5 & Climate modeling, atmospheric chemistry \\
    P7 & Computer Science (Model Optimization) & PhD 4th year & AI-led & 5.0/5 & Machine learning optimization, algorithm design \\
    P8 & Speech Recognition (Hot Word Detection) & PhD 2nd year & AI-led & 4.0/5 & Neural networks, voice processing systems \\
    P9 & Philosophy & 10 years (Professor) & AI-led & 3.5/5 & Ethics, philosophy of mind, consciousness studies \\
    P10 & Image Processing (Tamper Detection) & 2 years (Postdoc) & AI-led & 4.8/5 & Computer vision, digital forensics, deep learning \\
    \bottomrule
    \end{tabular}
\end{table}
\newpage

\section{Mimitalk Framework Code}
\label{sec:mimitalk_framework_code}
The demonstration code for the MimiTalk framework is publicly available on GitHub at \url{https://github.com/LFM097384/MimiTalk.demo}. This repository contains example scripts for the main functional modules as well as detailed usage instructions, enabling users to quickly understand and reproduce the key implementation processes described in this study. For further technical details or related questions, please refer to the documentation provided in the repository.

\end{document}